\documentclass[traditabstract]{aa} 
\usepackage{graphicx}
\usepackage{txfonts}
\usepackage{natbib}
\usepackage{longtable}
\usepackage{amsmath}

\bibpunct[]{(}{)}{;}{a}{}{,}

\begin{document}

   \title{Rotational and rovibrational spectroscopy of CD$_3$OH with an account of 
          CD$_3$OH toward IRAS 16293$-$2422\thanks{Transition 
          frequencies from this and earlier work are given as supplementary material. 
          We  also provide quantum numbers, uncertainties, and residuals between 
          measured frequencies and those calculated from the final set of 
          spectroscopic parameters. The data are available at Centre de Données astronomiques de
          Strasbourg (CDS) via anonymous 
          ftp to cdsarc.u-strasbg.fr (130.79.128.5) or via 
          http://cdsweb.u-strasbg.fr/cgi-bin/qcat?J/A+A/}}

   \author{
           V.~V. Ilyushin\inst{1,2}
           \and
           H.~S.~P. M{\"u}ller\inst{3}
           \and
           J.~K. J{\o}rgensen\inst{4}
           \and
           S. Bauerecker\inst{5}
           \and
           C. Maul\inst{5}
           \and
           Y. Bakhmat\inst{1}
           \and
           E.~A. Alekseev\inst{1,2}
           \and
           O. Dorovskaya\inst{1}
           \and
           S. Vlasenko\inst{6}
           \and
           F. Lewen\inst{3}
           \and
           S. Schlemmer\inst{3}
           \and
           K. Berezkin\inst{5,7}
           \and
           R.~M. Lees\inst{8}
           }

   \institute{Institute of Radio Astronomy of NASU, Mystetstv 4, 61002 Kharkiv, Ukraine\\
              \email{ilyushin@rian.kharkov.ua}
              \and
              Quantum Radiophysics Department, V.~N. Karazin Kharkiv National University, 
              Kharkiv, Ukraine
              \and
              I.~Physikalisches Institut, Universit{\"a}t zu K{\"o}ln,
              Z{\"u}lpicher Str. 77, 50937 K{\"o}ln, Germany\\
              \email{hspm@ph1.uni-koeln.de}
              \and
              Niels Bohr Institute, University of Copenhagen, {\O}ster Voldgade 5$-$7, 1350 Copenhagen K, Denmark 
              \and
              Institut f{\"u}r Physikalische und Theoretische Chemie, 
              Technische Universit{\"a}t Braunschweig, Gau{\ss}str. 17, 38106 Braunschweig, Germany
              \and
              Organic Chemistry Department, School of Chemistry, 
              V.~N. Karazin Kharkiv National University, Kharkiv, Ukraine              \and
              National Research Tomsk Polytechnic University, Tomsk 634050, Russia 
              \and
              Department of Physics, University of New Brunswick, Saint John, NB E2L 4L5, Canada
              }

   \date{Received XX YYY 2021 / Accepted ZZ WWW 2021}
 
\abstract

\abstract{Solar-type protostars have been shown to harbor highly deuterated complex organics, as evidenced, for instance, by the high relative abundances of doubly and triply deuterated isotopologs. While this degree of deuteration may provide important clues in studying the formation of these species, spectroscopic information on multiply deuterated isotopologs is often insufficient. In particular, searches for triply deuterated methanol, CD$_3$OH, are hampered to a large extent by the lack of intensity information from a spectroscopic model. The aim of the present study is to develop a spectroscopic model of CD$_3$OH in low-lying torsional states that is sufficiently accurate 
to facilitate further searches for CD$_3$OH in space. We performed a new measurement campaign 
for CD$_3$OH involving three spectroscopic laboratories that covers the 34~GHz$-$1.1~THz and the 
20$-$900~cm$^{-1}$ ranges. The analysis was performed using the torsion-rotation 
Hamiltonian model based on the rho-axis
method. We determined a model that describes the ground and first excited torsional 
states of CD$_3$OH, up to quantum numbers $J \leqslant 55$ and $K_a \leqslant 23$, and we derived 
a line list for radio-astronomical observations. The resulting line list is accurate up to 
at least 1.1~THz and should be sufficient for all types of radio-astronomical searches 
for this methanol isotopolog. This line list was used to search for CD$_3$OH in data from 
the Protostellar Interferometric Line Survey of IRAS 16293$-$2422 using the Atacama Large 
Millimeter/submillimeter Array. Specifically, CD$_3$OH is securely detected in the data, with a large number 
of clearly separated and well-reproduced lines. We not only detected lines belonging to 
the ground torsional state, but also several belonging to the first excited torsional state. The derived column density of CD$_3$OH and abundance relative to non-deuterated isotopolog confirm the significant enhancement of this multiply deuterated variant. This finding is in line with other observations of multiply deuterated complex organic molecules and may serve as an important constraint on their formation models.}


\authorrunning{V.~V. Ilyushin et al.}
\titlerunning{Spectroscopy of CD$_3$OH}

\maketitle
\hyphenation{For-schungs-ge-mein-schaft}

\section{Introduction}
\label{intro}

Methanol, CH$_3$OH, is among the very first molecules detected in space by means 
of radio astronomy \citep{det_CH3OH_1970}. It is among the most abundant polyatomic molecules in the interstellar medium and is observed both in its solid state and gas phase toward star forming regions \citep[e.g.,][]{Herbst2009}. It is an important product of the chemistry occurring on the icy surfaces of dust grains \citep[e.g.,][]{Tielens1982, Garrod2006} and often taken as a reference for studies of the chemistry of more complex organic molecules \citep[e.g.,][]{Jorgensen2020}. Furthermore, as a slightly asymmetric rotor, whose excitation is strongly dependent on kinetic temperature,  methanol  presents a useful diagnostic tool for estimating the conditions present in star-forming regions \citep{Leurini2004}. In fact, CH$_3$OH emission has been used to characterize physical conditions, for example, in Orion~KL \citep{MeOH_HIFI_Orion-KL_2011}, Sagittarius~B2(N2) \citep{ROH_RSH_2016}, and the solar-type protostellar system IRAS 16293$-$2422 \citep{PILS_2016}.

The enrichment of deuterium in dense molecular clouds has been known for many years 
\citep{Jefferts1973, Wilson1973}, and explanations have been proposed nearly as far back as this \citep{deuteration_1989}. The singly deuterated methanol isotopomers, CH$_3$OD \citep{det_CH3OD_1988} and CH$_2$DOH \citep{det_CH2DOH_1993}, were detected first. Some time later, \citet{det_CHD2OH_2002} 
observed CHD$_2$OH toward IRAS 16293$-$2422, followed by \citet{det_CD3OH_2004} detecting 
CD$_3$OH toward the same object. In addition, CHD$_2$OH was also found toward the low-mass protostars 
IRAS~2, IRAS~4A, and IRAS~4B in the NGC~1333 molecular cloud \citep{more_CHD2OH_2006}.

The degree of deuteration has been considered an indicator of the evolution of 
low-mass star-forming regions \citep{deuteration_2005,deuteration_2007,deuteration_2018}.
High degrees of deuteration were found in hot corinos, which are the warm and dense parts of 
low-mass star-forming regions, such as IRAS 16293$-$2422B \citep{deuteration_16293_2018}, 
the denser parts of the envelopes of low-mass star-forming regions, such as L483 
\citep{L483_3mm-survey_2019}, and even starless cores, such as TMC-1 
\citep{DC7N_TMC-1_2018,DCCNC_DNC3_TMC-1_2020,HDCCN_TMC-1_2021} and L1544 \citep{c-C3D2_L1544_TMC-1C_2013}. 
However, the deuterium enrichment is less pronounced in high-mass star-forming regions, 
such as NGC~6334I \citep{deuteration_NGC6334_2018}, and even less if they reside 
in the Galactic center, such as Sagittarius~B2(N2) \citep{EMoCA_with-D_2016}.

The high sensitivity of the Atacama Large Millimeter/submillimeter Array (ALMA) has made 
it possible to characterize the deuterium fractionation of various complex organic 
molecules comprehensively, for example, revealing some systematic differences between 
groups of species \citep[e.g.,][]{deuteration_16293_2018}. 
In particular, the data have made it possible to shed new light on the presence 
of multiply-deuterated variants \citep{H2CO_PILS_2018} and allowed, for the first time, 
the identifications of doubly deuterated  organics: methyl  cyanide (CHD$_2$CN; 
\citealt{CH3CN_PILS_2018}), methyl formate (CHD$_2$OCHO; \citealt{D2-MeFo_2019}), 
and dimethyl ether (CHD$_2$OCH$_3$; \citealt{D2-DME_2021}) toward the low-mass protostellar 
system IRAS 16293$-$2422. These multi-deuterated variants appear overabundant compared to 
the D/H ratio inferred from the singly and non-deuterated species, similar to what has 
previously been inferred, and which may reflect their formation processes at low temperatures 
\citep{deuteration_in_ice_2014}. These multiply deuterated variants, with their high abundances, 
may be responsible for many prominent lines that are observable with ALMA toward solar-type protostars.

The rotational spectrum of CD$_3$OH was studied already in the early days of microwave 
spectroscopy in the context of other methanol isotopologs, in particular, in order to 
determine the molecular structure \citep{MeOH-isos_1-0_I_1955,MeOH-isos_1-0_II_1956}. 
\citet{CH_D3OH_D_rot_1968} published the first extensive study of its rotational 
spectrum in the millimeter wave region by investigating the torsion-rotation interaction 
in the methanol isotopologs CH$_3$OH, CD$_3$OH, and CH$_3$OD. Additional accounts on 
its rotational spectrum were published with infrared \citep{CD3OH_FIR_1993} and 
microwave accuracy \citep{3isos_rot_1992,CD3OH_rot_1993,CD3OH_mmW_1998}. The latest of these studies also led to a thorough analysis of the CD$_3$OH rotational 
spectrum in the ground and first excited torsional states \citep{CD3OH_fitting_1998}. In that study, the authors analyzed a dataset consisting of 472 microwave and 5320 
far-infrared lines corresponding to transitions with $J$ up to 20 and $K$ up to 15 in the 
ground and first excited torsional states. The microwave part of the dataset 
covered the frequency range from 8 to 442~GHz, where all lines were assigned an uncertainty 
of 100~kHz or larger. The rovibrational spectrum of CD$_3$OH was also examined in several studies, with \citet{CD3OH_rho-ip_2015} being the most recent one to specifically deal with the in-plane rocking vibration.

The goal of our current study is to develop a spectroscopic model of CD$_3$OH in low-lying 
torsional states sufficiently accurate to provide reliable calculations for astronomical searches 
for CD$_3$OH in the interstellar medium. New measurements were carried out in broad frequency 
ranges extending the microwave dataset up to 1.1~THz. New far infrared measurements were carried 
out in addition in order to obtain more information on the torsional fundamental band. 
The rotational quantum number range coverage was extended to $J$ = 55 and $K$ = 23, and a fit 
within experimental uncertainties was obtained for the ground and first excited torsional states 
of the CD$_3$OH molecule using the so-called rho-axis-method. We searched for CD$_3$OH in the ALMA 
data of the Protostellar Interferometric Line Survey \citep[PILS,][]{PILS_2016} of the deeply embedded protostar IRAS 16293$-$2422 employing calculations of the rotational spectrum based on our results. CD$_3$OH is confidently detected with a large number of unblended or slightly blended and well-reproduced lines.

The rest of the manuscript is organized as follows. Section~\ref{exptl} provides details on 
our laboratory measurements. The theoretical model, spectroscopic analysis, 
and fitting results are presented in Sections~\ref{spec_backgr} and \ref{lab-results}. 
Section~\ref{astrosearch} describes our astronomical observations and the results of our search 
for CD$_3$OH, while Section~\ref{conclusion} gives the conclusions of our investigation.

\section{Experimental details}
\label{exptl}


\begin{figure}
\centering
\includegraphics[width=8cm,angle=0]{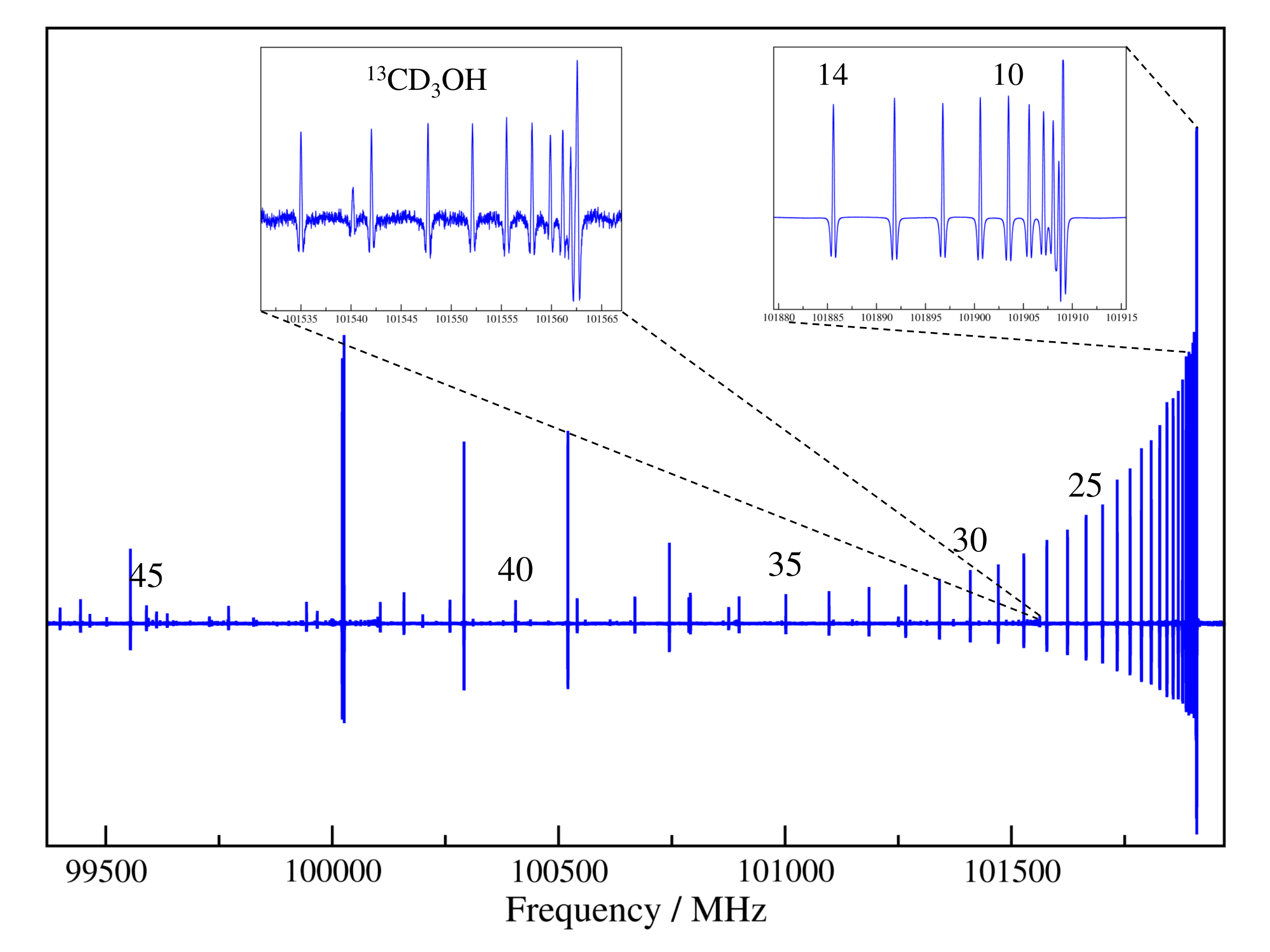}

\caption{Section of the millimeter spectrum of CD$_3$OH displaying part of the $\varv _{\rm t} = 1$, 
         $K = 0 - 1$ $b$-type $Q$-branch of E symmetry. Selected $J$ quantum numbers are given 
         above the lines. The right insert shows the origin on an expanded scale. The left insert 
         depicts part of the corresponding branch of $^{13}$CD$_3$OH with natural $^{12}$C/$^{13}$C ratio.}
\label{fig_3mm_Cologne}
\end{figure}

\subsection{Rotational spectra at IRA NASU}

The measurements at the Institute of Radio Astronomy (IRA) of NASU were performed in the 
frequency ranges 34$-$183~GHz and 354$-$416~GHz using an automated millimeter wave spectrometer 
\citep{Alekseev2012}, built according to the so-called "classical" scheme of absorption 
spectrometers. The synthesis of the frequencies in the millimeter wave range is carried out by a 
two-step frequency multiplication of a reference synthesizer in two phase-lock-loop stages. 
At the second step of frequency multiplication, a set of backward wave oscillators is used 
to cover the frequency range from 34 to 183~GHz. Additional application of a solid state tripler 
from Virginia Diodes, Inc. (VDI) extended our measurements to the 354$-$416 GHz range. 
We used a commercial sample of CD$_3$OH and all measurements were carried out at room temperature 
with sample pressures providing linewidths close to the Doppler limited resolution (about 2~Pa). 
Estimated uncertainties were 10~kHz, 30~kHz, and 100~kHz depending on the observed signal-to-noise 
ratios (S/N).

\subsection{Rotational spectra at the Universit{\"at} zu K{\"o}ln}

The measurements at the Universit{\"a}t zu K{\"o}ln were recorded at room temperature 
using three different spectrometers. Pyrex glass cells of different lengths and with 
an inner diameter of 100~mm were employed. The cells were equipped with Teflon windows 
below 510~GHz, whereas high-density polyethylene was used at higher frequencies. All spectrometer systems used VDI frequency multipliers 
driven by Rohde \& Schwarz SMF~100A microwave synthesizers as sources. Schottky diode 
detectors were utilized below 510~GHz, whereas a closed cycle liquid He-cooled InSb 
bolometer (QMC Instruments Ltd) was applied between 760 and $\sim$1094~GHz.
Frequency modulation was used throughout. The demodulation at $2f$ causes an isolated 
line to appear close to a second derivative of a Gaussian.

Two connected cells, each of 7~m in length,  were used in a double pass arrangement at pressures of 2~Pa 
to cover 70$-$120~GHz; \citet{n-BuCN_rot_2012} provide additional details on this spectrometer. 
Figure~\ref{fig_3mm_Cologne} illustrates the sensitivity of this spectrometer. Alongside the $Q$-branch 
of CD$_3$OH in the first excited torsional state, part of the same branch is shown for 
$^{13}$CD$_3$OH in natural $^{13}$C/$^{12}$C ratio.

A  double pass cell of 5 m in length was used to cover 150$-$251~GHz, 249$-$375~GHz, and 375$-$505~GHz. 
The pressures were in the 2~Pa range for the 150$-$375~GHz regions and 1.5~Pa at 375$-$505~GHz. 
Further information on this spectrometer is available elsewhere \citep{OSSO_rot_2015}. 
We achieved frequency accuracies of 5~kHz with both spectrometers in a study of 2-cyanobutane 
\citep{2-CAB_rot_2017} with a much richer rotational spectrum. 
We employed a setup with a 5~m single pass cell to cover 760~GHz to about 1094~GHz 
at pressures of $\sim$1.5~Pa. Our studies on isotopic formaldehyde \citep{H2CO_rot_2017} 
or thioformaldehyde \citep{H2CS_rot_2019} demonstrate that accuracies of 10~kHz can be 
reached readily routinely for very symmetric lines with good S/N. We assigned uncertainties 
of 10~kHz for the most precise measurements in our fits to accomodate possible measurement 
errors due to unresolved hyperfine structure caused by the nonzero electric quadrupole moment 
of the deuterium atoms; these errors may also be due to possible saturation distortions for strong lines. 
Other lines were given uncertainties of 20~kHz, 30~kHz, 50~kHz, 100~kHz, and 200~kHz, 
depending on the observed S/N and on the frequency range.

\subsection{Far-infrared spectra at the Technische Universit{\"at} Braunschweig}

Measurements at the Technische Universit{\"a}t Braunschweig were performed with the 
Z{\"u}rich prototype Fourier transform infrared spectrometer Bruker IFS125HR \citep{Albert2011}, 
which was slightly optimized to obtain the maximum nominal resolution for a nine-chamber system 
of 0.00096~cm$^{-1}$. The resolution is defined as 1/$d_{\rm MOPD}$, with $d_{\rm MOPD}$ 
standing for the maximum optical path difference \citep{Albert2011}. 
Two pairs of records have been obtained for CD$_3$OH using a stainless steel multireflection White cell at four paths, giving a total optical pathlength of 4.05 m. The 1.6~K Si bolometer of IRLabs was used 
in the spectral region of 20–700~cm$^{-1}$, with a globar as radiation source, a 3.5~$\mu$m Mylar beam–splitter, and Teflon windows. A resolution of 0.003cm$^{-1}$ was applied at sample pressures 
of 0.195~mbar and 1.0~mbar. A liquid–helium cooled germanium–copper (Ge:Cu) detector was applied in 
the spectral region of 350–900 cm$^{-1}$. This detector was combined with an optical filter 
which is also liquid–helium cooled to optimize the noise in this spectral region. 
A globar radiation source, KBr beamsplitter, and CsI windows  were used for recording 
the spectra of CD$_3$OH with 0.00096~cm$^{-1}$ resolution at sample pressures of 0.5~mbar and 1.0~mbar. In the current study, the 350–900 cm$^{-1}$ record obtained at the sample pressure of 1.0~mbar and the 20–700~cm$^{-1}$ record obtained at the sample pressure of 0.195~mbar were used.     
Both these spectra were calibrated with residual water lines, whose frequencies were taken from the HITRAN database \citep{GORDON2017}. Uncertainties of 0.0004 cm$^{-1}$ were applied to the far-infrared (FIR) 
transition frequencies.

\section{Spectroscopic properties of CD$_3$OH and our theoretical approach}
\label{spec_backgr}

As one of the lightest and structurally simplest molecules that is capable of internal rotation, 
methanol may be designated, from a spectroscopic point of view, as the prime test case for theoretical 
models treating large amplitude torsional motion in molecules. Methanol is a nearly prolate top 
($\kappa \approx -0.982$) that is characterized by a rather high coupling between internal 
and overall rotations in the molecule ($\rho \approx 0.81$). In the case of the triply deuterated 
isotopolog CD$_3$OH, the asymmetry parameter is nearly the same ($\kappa \approx -0.977$), 
whereas the coupling is slightly higher ($\rho \approx 0.89$). The torsional potential barrier 
$V_3$ is about 370~cm$^{-1}$ for both CH$_3$OH and CD$_3$OH. The torsional splittings can 
reach tens of gigahertz even in the torsional ground state. The torsional problem may be 
considered as an intermediate barrier case \citep{RevModPhys.31.841} since $s$ is $\sim$6.0 
for CH$_3$OH and $\sim$6.6 for CD$_3$OH; here $s = 4V_3/9F$ is the reduced barrier, where $F$ is the rotation constant of the internal rotor. 
The torsional effects in CH$_3$OH and CD$_3$OH should be quite similar because of 
the similar values of $s$. However, we expect that rotational levels with higher $J$ and $K$ 
values will be accessible in a room temperature experiment for CD$_3$OH compared to CH$_3$OH 
because of the much smaller rotational parameters, $A \approx 2.36$~cm$^{-1}$, 
$B \approx 0.662$~cm$^{-1}$, $C \approx 0.642$~cm$^{-1}$ in CD$_3$OH versus 
$A \approx 4.25$~cm$^{-1}$, $B \approx 0.823$~cm$^{-1}$, $C \approx 0.792$~cm$^{-1}$ in 
CH$_3$OH \citep{XU:2008305}.

In the current study, we aim to extend the previous analysis to the higher $J$ levels, 
above the previously achieved limit of $J \leq 20$ \citep{CD3OH_fitting_1998}, 
for the lowest torsional states of CD$_3$OH. An analogous task was successfully fulfilled 
in the past for similar molecules such as acetaldehyde \citep[CH$_3$CHO;][]{Smirnov:2014} with 
$\rho \approx 0.33$ and methyl mercaptan \citep[CH$_3$SH;][]{Zakharenko:2019}, with $\rho \approx 0.65$, 
where a high level of coupling between internal and overall rotations poses significant 
problems in treating torsion-rotation spectra. In particular, CD$_3$OH, a molecule in which the coupling is further increased ($\rho \approx 0.89$), represents a good testing case for available theoretical 
approaches and computer codes implementing those approaches.

The approach in the present study is the so-called rho-axis-method (RAM), which has proven 
to be the most effective approach so far in treating torsional large amplitude motions in 
methanol-like molecules. The method is based on the work of \citet{Kirtman:1962}, 
\citet{CH_D3OH_D_rot_1968}, and \citet{Herbst:1984}. It takes its name from the choice 
of the axis system \citep{Hougen:1994}, namely, the rho axis system, which is related 
to the principal axis system $a$, $b$, $c$ by a rotation chosen to eliminate the 
$-2F\rho_{x}p_{\alpha}J_{x}$ and $-2F\rho_{y}p_{\alpha}J_{y}$ coupling terms in the 
kinetic energy operator; here $F$ is the internal rotation constant, $p_{\alpha}$ 
is the internal rotation angular momentum, $J_{x}$ and $J_{y}$ are the usual $x$ and $y$ 
components of the global rotation angular momentum, and $\rho$ is a vector that 
expresses the coupling between the angular momentum of the internal rotation $p_{\alpha}$ 
and that of the global rotation $J$. The rotation to the RAM axis system corresponds 
to making the new $z$ axis coincident with the $\rho$ vector, thus making 
$\rho_{x}=\rho_{y} = 0$ by definition. The angle between the RAM a-axis and the 
principal-axis-method (PAM) a-axis is only 0.07$^{\circ}$ and 0.14$^{\circ}$ in 
CH$_3$OH and CD$_3$OH, respectively. Such a small angle in combination with the low 
asymmetry ($\kappa \approx -0.98$) means that the RAM $a$-axis in methanol should be 
suitable for the $K$ quantization and that eigenvectors can be assigned unambiguously 
using the dominant basis set components.

We employed the RAM36 code \citep{Ilyushin:2010,Ilyushin:2013}, which was successfully used 
in the past for a number of near prolate tops with rather high $\rho$ and $J$ values, 
such as acetaldehyde with $\rho \approx 0.33$ and $J_{\rm max}=66$ \citep{Smirnov:2014}, 
methyl arsine with $\rho \approx 0.41$ and $J_{\rm max}=50$ \citep{Motiyenko:2020}, and 
methyl mercaptan with $\rho \approx 0.65$ and $J_{\rm max}=61$ \citep{Zakharenko:2019}. 
The code carries out the RAM approach for molecules with a $C_{3{\rm v}}$ top attached to 
a molecular frame of $C_s$ or $C_{2{\rm v}}$ symmetry and having a threefold or sixfold barrier 
to internal rotation, respectively. The RAM36 code uses the two-step diagonalization 
procedure of \citet{Herbst:1984}. In the current study, we keep 31 torsional basis 
functions at the first diagonalization step and 11 torsional basis functions at the 
second diagonalization step. The labeling scheme after the second diagonalization step 
is based on an eigenfunction composition by searching for a dominant eigenvector component. 
The energy levels are labeled by the free rotor quantum number, $m$, the overall rotational 
angular momentum quantum number, $J$, and a signed value of $K_a$, which is the axial 
$a$-component of the overall rotational angular momentum $J$. In the case of the A 
symmetry species, the $+/-$ sign corresponds to the so-called "parity" designation, 
which is related to the A1/A2 symmetry species in the group $G_6$ \citep{Hougen:1994}. 
The signed value of $K_a$ for the E symmetry species reflects the fact that the 
Coriolis-type interaction between the internal rotation and the global rotation 
causes the $|K_a| > 0$ levels to split into a $K_a > 0$ level and a $K_a < 0$ level.

\section{Spectroscopic results}
\label{lab-results}

We started our analysis from the results of \citet{CD3OH_fitting_1998}. The dataset, consisting of 
472 microwave and 5320 FIR transitions, ranging up to $\varv_{\rm t} = 1$, $J_{\rm max} = 20$, and 
$K_{\rm max} = 15$, was fit with 54 parameters of the RAM Hamiltonian, and a weighted standard 
deviation of 0.966 was achieved \citep{CD3OH_fitting_1998}. Unfortunately, we were able to recover 
only the microwave part of the dataset from this paper, which was presented in Tables 3a and 3b of 
\citet{CD3OH_fitting_1998}. We refit the available microwave part of the CD$_3$OH dataset with the 
RAM36 program \citep{Ilyushin:2010,Ilyushin:2013} as the first step; the BELGI code \citep{Kleiner:2010} 
was used in the previous study \citep{CD3OH_fitting_1998}. The resulting fit was the starting point 
of our present investigation.


\begin{figure}
\centering
\includegraphics[width=8cm,angle=0]{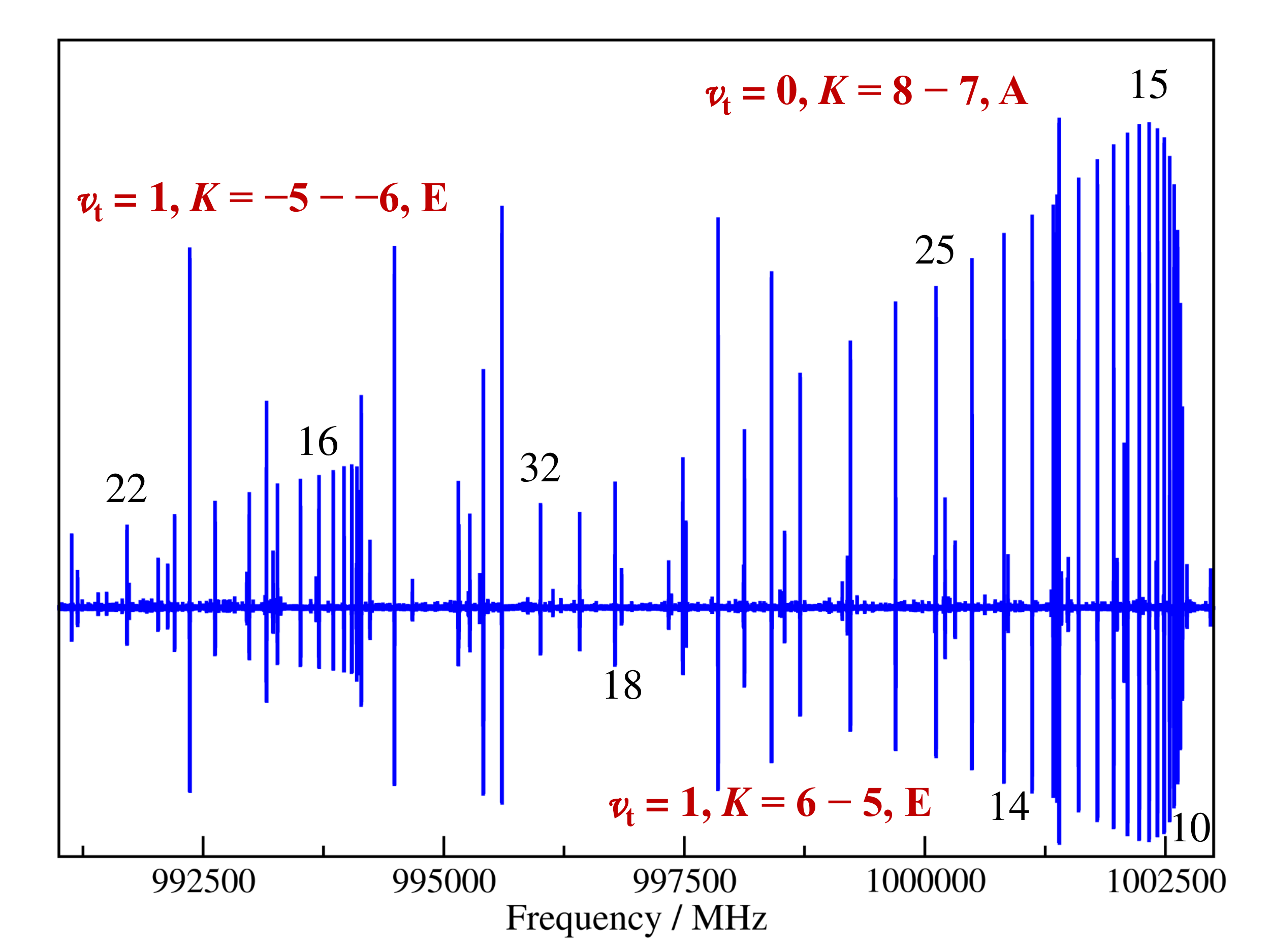}

\caption{Section of the submillimeter spectrum of CD$_3$OH. Parts of two narrow $b$-type $Q$-branches 
   are presented with designations and selected $J$ quantum numbers above the lines. Also shown is a 
   weaker and wider $Q$-branch with designations and selected $J$ quantum numbers below the lines.}
\label{fig_THz_Cologne}
\end{figure}


New data were assigned starting from the Kharkiv measurements in the 34–183~GHz frequency range. 
Transitions with $J$ up to 49 have been assigned at this stage and the first transitions belonging 
to the second excited torsional state of CD$_3$OH were assigned. The FIR records were analyzed 
subsequently on the basis of our new results. Submillimeter wave and THz measurements from Cologne 
and Kharkiv were assigned at the same time. The CD$_3$OH THz spectrum is rather dense in parts, 
as can be seen in Fig.~\ref{fig_THz_Cologne}. The assignments for the three torsional states, namely: 
$\varv_{\rm t} = 0$, 1, and 2, were done in parallel. Whenever it was possible, we replaced the 
old measurements from \citet{CD3OH_fitting_1998} with the more accurate new ones. 
At the same time, we decided to keep in the fits two measured values for the same transition from the 
Kharkiv and Cologne spectral recordings in those parts of the frequency range where the measurements 
from the two laboratories overlap (150$-$183~GHz and 354$-$416~GHz). We emphasize that a rather good 
agreement within the experimental uncertainties was observed for this limited set of duplicate measurements.


\begin{figure}
\centering
\includegraphics[width=9cm,angle=0]{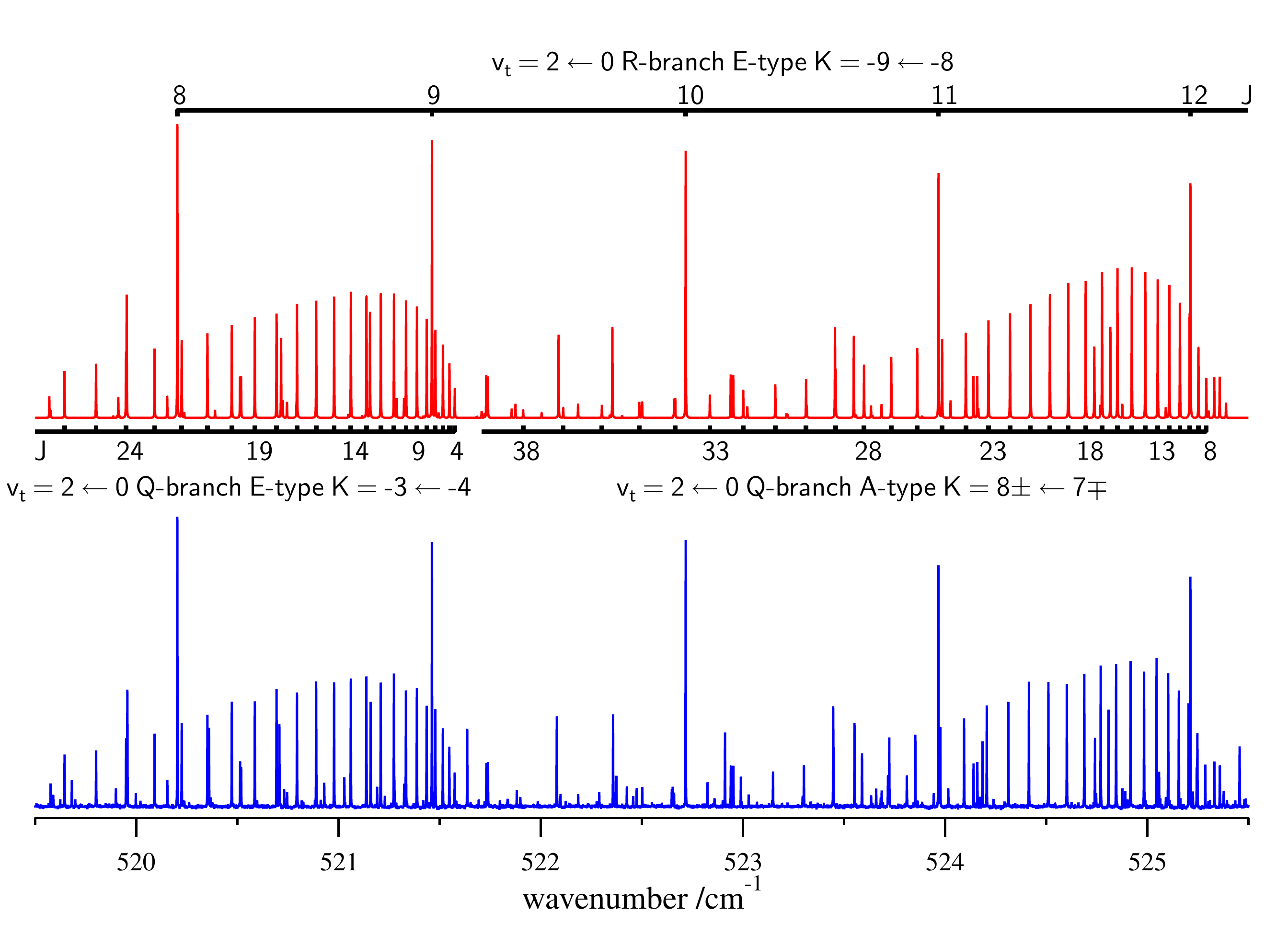}

\caption{Portion of the CD$_3$OH FIR spectrum dominated by one $R$-branch and two $Q$-branches 
  of the $\varv_{\rm t} = 2 \leftarrow 0$ band in the 519.5$-$525.5~cm$^{-1}$ range. The observed 
  spectrum is shown in the lower panel and the calculated one in the upper panel. The experimental 
  frequencies and the  intensity pattern are rather well reproduced by our model for the spectral 
  features dominating this frequency range. Assignments of lower state $J$ quantum numbers are 
  given at the top and the bottom of the upper panel for these branches.}\label{fig_FIR}
\end{figure}


A search of the optimal set of RAM torsion–rotation parameters was carried out in parallel with 
the assignment process by testing different parameters up to an order of $n_{\rm op} = 12$; here, we employed 
the ordering scheme of \citet{Nakagawa:1987}. In the process of model refinement, it became evident 
rather soon that the $\varv_{\rm t}=2$ torsional state poses significant problems with fitting. 
Although our understanding of the $\varv_{\rm t}=2$ torsional state was qualitatively rather good (see Fig.~\ref{fig_FIR}, for example, comparing simulated and observed spectra for a part of the $\varv_{\rm t}=2 \leftarrow 0$ torsional band), the best weighted root mean square deviation we could obtain for a combined fit of the $\varv_{\rm t}=0$ to 2 CD$_3$OH states with $J$ up to 54 was as large as 4.4. The strong influence of 
intervibrational interactions arising from low lying small amplitude vibrations in the molecule, 
which then propagate down through numerous intertorsional interactions, is a possible explanation 
for these problems. \citet{Weber1982} were the first to point out the possibility of Fermi resonances with high torsional levels for all methanol isotopologs, and a more recent overview of possible interactions with low lying small amplitude vibrations in CD$_3$OH may be found in \citet{CD3OH_vs_CH3OH_2020}. It is likely that the perturbations to $\varv_{\rm t}=2$  levels could be on the order of a few hundredths of a cm$^{-1}$, which is very substantial in terms of the fitting. Taking into account the astrophysical significance 
of methanol and its isotopologs, we decided subsequently to limit our fitting attempts mainly to 
the ground and first excited torsional states. Only the lowest three $K$ series for the A and E 
species in $\varv_{\rm t}=2$ were retained in order to constrain the torsional parameters in the 
Hamiltonian model better. These $\varv_{\rm t}=2$ $K$ levels are affected least by the intervibrational 
interactions arising from low lying small amplitude vibrations. This corresponds to $K = -4, -1, 2$ for the E species in $\varv_{\rm t}=2$ and to $K = -3, 0, 3$ 
for the A species. We concentrated on assigning high $K$ transitions in $\varv_{\rm t}=0$ and 1 
at the final stage of our analysis.

The final dataset in this work involves 7272 microwave and 5619 FIR line frequencies, which, as a result of blending, correspond to 15798 transitions with $J_{\rm max}$ = 55. A fit with a weighted 
root mean square (rms) deviation of 0.81 with 117 parameters included in the model (one fixed 
parameter) was chosen as our “best fit” for this paper. The quality of this fit can be seen 
in Table~\ref{tbl:statisticInf}. The overall weighted rms deviation of 0.81 and the additional 
fact that all data groups are fit within experimental uncertainties seems satisfactory to us (see also the left part of Table~\ref{tbl:statisticInf} where the data are grouped by 
measurement uncertainty). The weighted rms deviations for the data grouped by torsional state also 
demonstrate a rather good agreement between our model and the experiment. It should be noted that absolute calibration of the FIR spectra against the frequencies of residual water impurity lines appeared to be important in achieving a fit within experimental error, especially for the most precise group of measurements with 10 kHz measurement uncertainty (at the initial stage of our model refinement, the FIR frequencies, obtained employing the original Bruker calibration, were used for the fitting). The 117 molecular parameters used in our final fit are given in Table~\ref{tbl:ParametersTable}. 
The numbers of the terms in the model distributed between the orders $n_{\rm op} = 2$, 4, 
6, 8, 10, 12 are 7, 22, 40, 34, 13, 1, respectively. This is consistent with the limits of 
determinable parameters of 7, 22, 50, 95, 161, and 252 for these orders, as calculated 
from the differences between the total number of symmetry-allowed Hamiltonian terms of order 
$n_{\rm op}$ and the number of symmetry-allowed contact transformation terms of order 
$n_{\rm op} - 1$ (again we apply here the ordering scheme of \citet{Nakagawa:1987}).

We decided to fix the $V_9$ parameter in the final fit to improve the convergence of the fit. 
The fit with $V_9$ varied was still converging with respect to the relative change in the 
weighted rms deviation of the fit at the last iteration, but the corrections to some of 
the parameters values were as large as $\sim$10$^{-1}$ of their confidence intervals in 
the last iteration and concomitant changes in the calculated transition frequencies 
exceeded 1~kHz for some transitions. However, we generally expect the ratio usually to be about 
$10^{-3}$ to $10^{-4}$ for a "good" convergence, and changes in the calculated transition 
frequencies are usually below 1~kHz even for FIR data. The $V_9$ parameter is larger 
than $V_6$ in magnitude, as shown in Table~\ref{tbl:ParametersTable}, indicating that the 
expansion of the torsional potential function is far from a smooth convergence. 
As was discussed previously for methyl mercaptan \citep{Zakharenko:2019}, such behavior 
of the torsional potential function may be attributed to intervibrational interactions 
with low-lying small amplitude vibrational modes in the molecule. This may also be a reason 
for our convergence problems if $V_9$ is varied because the fit attempts to account for 
perturbations by varying a number of parameters, but the Hamiltonian terms needed to account explicitly for the perturbations are not yet built in our present model; this, in turn, leads to a high 
correlation among some parameters and $V_9$. Omitting $V_9$ from our model was not an option 
because this would lead to a significant increase in the weighted rms deviation of the fit. 
Therefore, we decided to keep $V_9$ fixed to a value we obtained from our last best fit with $V_9$ varied, and then truncated to the number of significant digits provided by that fit.      


\begin{figure}
\centering
\includegraphics[width=9cm,angle=0]{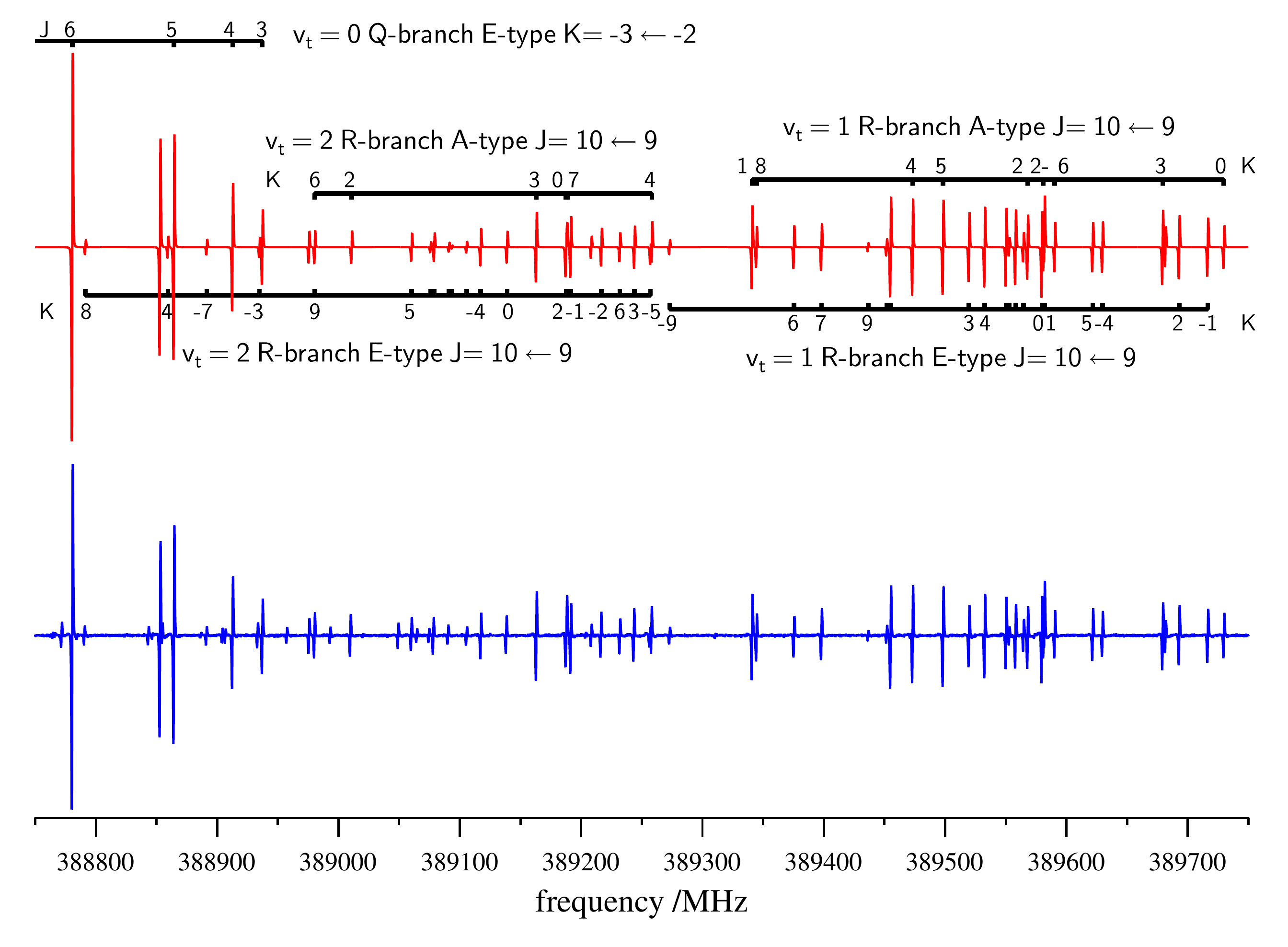}

\caption{Portion of the CD$_3$OH microwave spectrum dominated by one $\varv_{\rm t} = 0$ $Q$-branch and two $R$-branches from the $\varv_{\rm t} = 1,2$ excited torsional states in the 388.75$-$389.55~GHz range. The observed spectrum is shown in the lower panel and the calculated one in the upper panel. The experimental frequencies and the  intensity pattern are rather well reproduced by our model for the spectral features dominating this frequency range, although for some $\varv_{\rm t} = 2$ E type transitions (see e.g., $K=4, J = 10 \leftarrow 9$ transition near 388859 MHz), there are noticeable shifts in frequency with respect to our predictions. The $J$ and selected $K$ quantum numbers are given at the top and the bottom of the upper panel for the dominant spectral branches. We note that for A type transitions of the R branches the $K$ quantum number stands for a degenerate pair of $K{\pm}$ transitions in all cases except $K$=0,1,2,2- in the $\varv_{\rm t} = 1$ state and $K$=0 in the $\varv_{\rm t} = 2$ state.}
\label{fig_MMW_Kharkiv}
\end{figure}


We calculated a list of CD$_3$OH transitions in the ground and first excited torsional states 
for astronomical observations using the parameters of our final fit. We employed the dipole moment 
function of \citet{MEKHTIEV1999171} in our calculations, where the values for the permanent 
dipole moment components were replaced by $\mu_a = 0.9080(9)$~D and  $\mu_b = 1.4378(10)$~D. 
These values were determined for $^{13}$CD$_3$OH from Stark effect measurements 
\citep{MUKHOPADHYAY2015119}, but based on the similarity between $^{12}$CH$_3$OH \citep{MUKHOPADHYAY201551} 
and $^{13}$CH$_3$OH \citep{SASTRY1994374} dipole moment components, we believe that these are 
the best estimates available for $^{12}$CD$_3$OH.  The permanent dipole moment components were 
rotated from the principal axis system to the rho axis system of our Hamiltonian model. As seen from Fig.~\ref{fig_MMW_Kharkiv}, this choice of dipole moment components provides an opportunity to reproduce the observed intensity pattern of the CD$_3$OH spectrum rather well.
The list of CD$_3$OH transitions includes information on transition quantum numbers, 
transition frequencies, calculated uncertainties, lower state energies, and transition strengths.    
Since extrapolation beyond the quantum number coverage of a given experimental dataset becomes 
rapidly unreliable, especially in the case of molecules with large amplitude motions, we chose 
a torsional state limit of $\varv_{\rm t} \leq 1$ and rotational limits of $J \leq 60$ and 
$|K_{a}| \leq 24$. We label the torsion-rotation levels by the free rotor quantum number, $m$, 
the overall rotational angular momentum quantum number, $J$, and a signed value of $K_a$, 
as mentioned earlier in this work. We also provide $K_c$ values for convenience, but they are 
recalculated from the $J$ and $K_{a}$ values, $K_{c} = J - |K_{a}|$ for $K_{a} \geq 0$ 
and $K_{c} = J - |K_{a}| + 1$ for $K_{a} < 0$. The $m$ values of 0, $-$3 / 1, $-$2 
correspond to A/E transitions of the $\varv_{\rm t} = 0$ and 1 torsional states, respectively. 
The calculations were truncated above 1.3~THz. Additionally, we limited our calculations to 
transitions for which the uncertainties are less than 0.1~MHz. 
The lower-state energies are given 
in reference to the $K_{a} = 0$ A-type $\varv_{\rm t} = 0$ level, calculated to be 124.1663857823~cm$^{-1}$ above the bottom of the torsional potential well. 
In addition we provide the torsion-rotation part of the partition function $Q_{rt}$(T) 
of CD$_3$OH calculated from first principles, that is, via direct summation over the 
torsion-rotational states. The maximum $J$ value is 100 and $n_{\varv_{\rm t}} = 11$ torsional 
states were taken into account. The calculations, as well as the experimental line list from 
the present work, can be found in the online Supplementary material with this article.

\section{Astronomical search for CD$_3$OH}
\label{astrosearch}

The new spectroscopic calculations were used to search for lines of CD$_3$OH in data from 
the Protostellar Interferometric Line Survey (PILS). As described in \citet{PILS_2016}, 
the core of PILS is an unbiased survey of the Class~0 protostellar system IRAS~16293$-$2422 
using ALMA at frequencies between 329 and 363~GHz (the main atmospheric window of ALMA's Band 7). The data image the spectral lines toward the multiple sources in the region with 
an angular resolution of $\sim$0.5$''$ and a spectral resolution of $\sim$0.2~km~s$^{-1}$. 
Toward one specific region that is slightly offset from the ``B'' component of the system 
(IRAS16293B in the following), the lines are narrow, $\sim$1~km~s$^{-1}$, and 
with the sensitivity of the ALMA data $\sim$10,000 individual lines can be identified 
$-$ including many complex organic molecules and their isotopologs (see, e.g., 
\citealt{PILS_2016,deuteration_16293_2018,CH3CN_PILS_2018,composition_16293A_2020}). This includes a number of new detections toward solar-type protostars or in the interstellar medium overall.

\begin{figure}
\resizebox{\hsize}{!}{\includegraphics{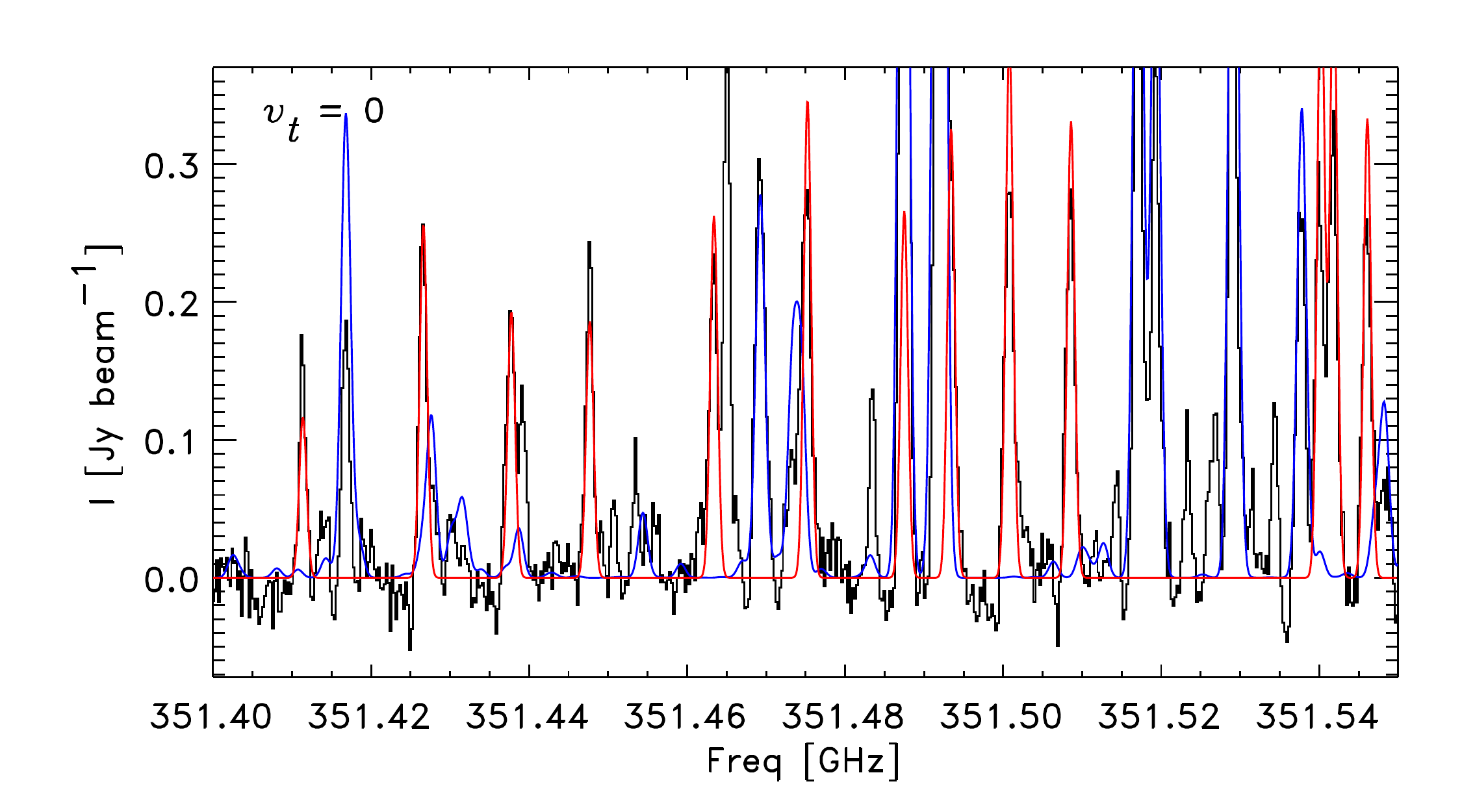}}
\resizebox{\hsize}{!}{\includegraphics{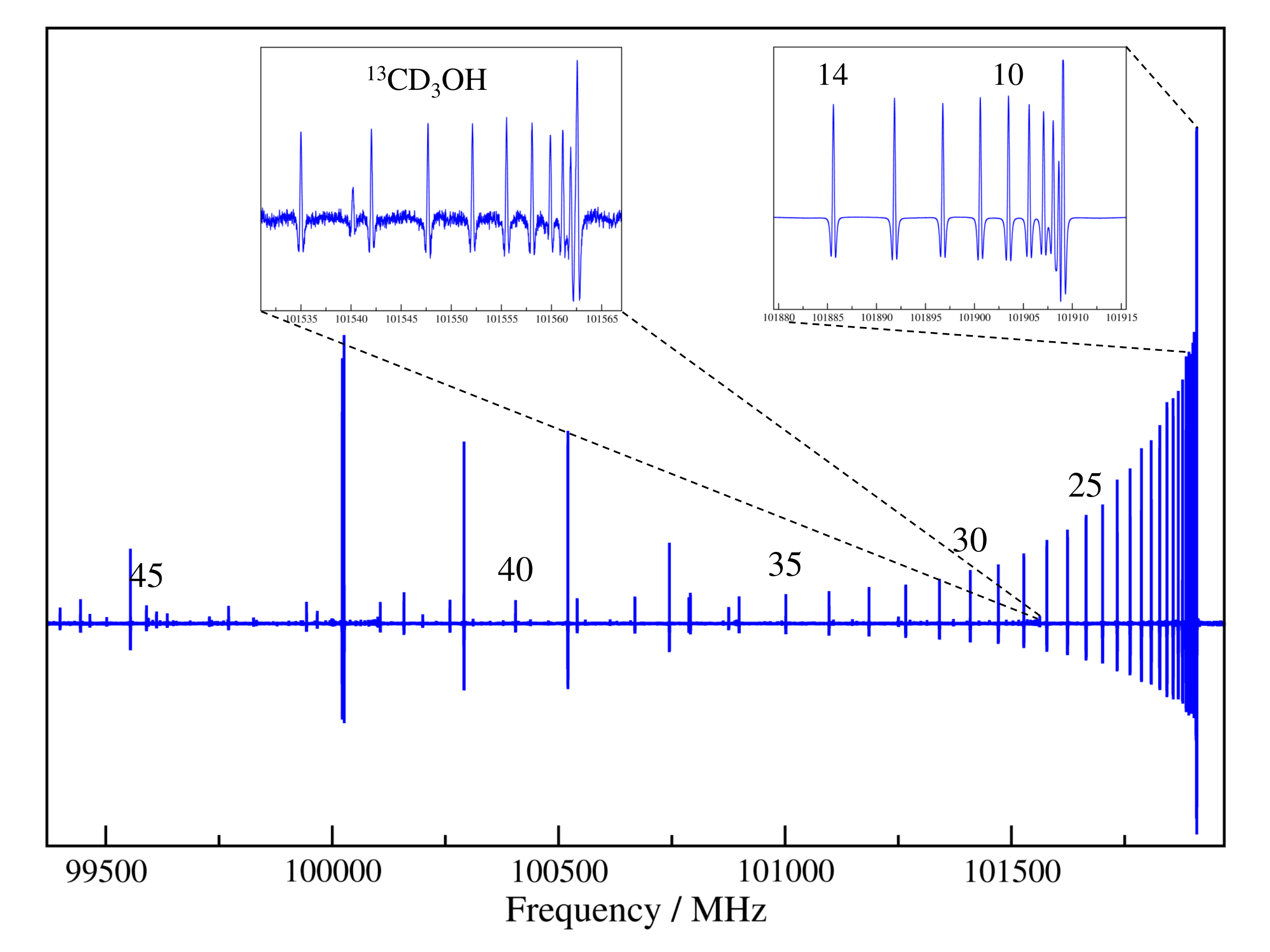}}
\caption{Example set of prominent $\varv_{\rm t} = 0$ (upper panel) and $\varv_{\rm t} = 1$ 
         (lower panel)  CD$_3$OH transitions in data of the ALMA PILS program 
         toward a position slightly offset from the ``B'' component of IRAS~16293$-$2422. 
         The black line shows the observed spectrum extracted toward the position, 
         the red line corresponds to the best fit synthetic spectrum model for the CD$_3$OH transitions 
         and the blue line presents the fits to other species identified within PILS.}
\label{astro-spectrumvt0vt1}
\end{figure}

The search was conducted in a similar manner as for other complex organics in the PILS 
survey by fitting synthetic spectra to the PILS data. Such synthetic spectra are calculated 
assuming that the excitation of the molecule is characterized by local thermodynamical 
equilibrium (LTE), which is reasonable at the densities on the scales probed by PILS 
\citep{PILS_2016}. Furthermore, a common velocity offset relative to the local standard 
of rest is adopted (2.6~km~s$^{-1}$) and the lines are fit with a similar width 
(1~km~s$^{-1}$; FWHM). We focus in the analysis on lines that are optically thin 
(fitted values of $\tau < 0.2$~km~s$^{-1}$), leaving the kinetic temperature and 
column density of the species as the two free parameters. Figure~\ref{astro-spectrumvt0vt1} (upper panel) shows an example of a fit to a prominent subset of $\varv_{\rm t} = 0$ lines at 351.5~GHz and 
Figs.~\ref{all_spec1}$-$\ref{all_spec6} in the appendix illustrate fits to the 48 lines calculated 
to be strongest while having $\tau < 0.2$. We also detected several unblended or only 
slightly blended transitions pertaining to the $\varv_{\rm t} = 1$ torsional state. Figure~\ref{astro-spectrumvt0vt1} (lower panel) shows an example of a fit to a prominent subset of $\varv_{\rm t} = 1$ lines at 350.6~GHz (see also Figs.~\ref{all_spec7}$-$\ref{all_spec9} in the appendix).
The best fit corresponds to an excitation temperature of 225~K for a CD$_3$OH 
column density of 3.1$\times 10^{16}$~cm$^{-2}$. 
Varying the excitation temperature from 200 to 250~K does not change the quality 
of the fits significantly and results in column densities of $2.8 \times 10^{16}$ 
to $3.2 \times 10^{16}$~cm$^{-2}$. The lowest excitation lines ($E_{\rm up} < 100$~K) may be tracing a potentially more extended, cold foreground layer \citep{deuteration_16293_2018}. If those lines are excluded, a reasonable fit can be obtained for temperatures up to 300~K with a column density of $3.7 \times 10^{16}$~cm$^{-2}$. CD$_3$OH is securely detected in the data at a 
high level of confidence with a number of clearly separated and well-reproduced lines. 
Furthermore, the strengths seen for a rather high number of transitions surpass many 
of the transitions of rarer species as well as their isotopologs 
\citep[see, e.g., examples in][]{deuteration_16293_2018}, highlighting the importance 
of the spectroscopic properties of the isotopic variants of molecules such as methanol.

We obtained a CD$_3$OH/CH$_3$OH ratio of  $\sim$0.33\% from this column density and 
that of CH$_3$OH, which was estimated from optically thin lines of CH$_3^{18}$OH 
\citep{PILS_2016,deuteration_16293_2018}. If all substitution reactions of deuterium 
in CH$_3$OH were equally probable \citep[e.g.,][]{det_CD3OH_2004,D2-MeFo_2019}, 
this CD$_3$OH/CH$_3$OH ratio would imply a D/H ratio of 15\%. 
This ratio is about a factor of 7 higher than the D/H ratios inferred from measuring 
the singly deuterated (CH$_2$DOH and CH$_3$OD) isotopologs relative to the main isotopolog 
but in line with the enhancements seen in the doubly deuterated variants of methyl formate (CH$_3$OHCHO; \citealt{D2-MeFo_2019}) and dimethyl ether 
(CH$_3$OCH$_3$; \citealt{D2-DME_2021}). Thus, this result lends further support 
to the idea that the enhanced D/H ratios of the multi-deuterated complex organics 
are inherited from the precursor molecules that are formed as the outer ice mantles 
of interstellar dust grains late in the protostellar evolution when the gas-phase [D]/[H] 
abundance is the highest \citep[e.g.,][]{deuteration_in_ice_2014,deuterated_CH3OH_2019}. 
However, it should be noted that the measured abundance ratios of CD$_3$OH relative 
to CH$_3$OH are still significantly above those predicted in the models by 
\citet{deuteration_in_ice_2014} of $\sim$0.01\%, while the abundances of the 
singly deuterated to non-deuterated methanol species were found to be in better 
agreement with model predictions \citep{deuteration_16293_2018,deuterated_CH3OH_2019}.

\section{Conclusion}
\label{conclusion}

In this work, we performed a new study of the torsion-rotation spectrum of CD$_3$OH isotopolog using 
a torsion-rotation RAM Hamiltonian. The new microwave measurements carried out in broad 
frequency ranges from 34~GHz to 1.1~THz were augmented by new FIR measurements. 
Transitions with $J$ up to 55 and $K_a$ up to 23 involving the $\varv_{\rm t}$ = 0, 1, 2 
torsional states were assigned and analyzed. After revealing perturbations in the second 
excited torsional state of CD$_3$OH, presumably caused by the intervibrational interactions 
arising from low-lying small-amplitude vibrations in this molecule, we concentrated our 
efforts on refining the theoretical model for the ground and first excited torsional states 
only. A fit within the experimental uncertainties (weighted rms deviation 0.81) was achieved 
for the dataset consisting of 7272 microwave and 5619 FIR line frequencies.

Calculations of the ground and first excited torsional states were carried out for 
astronomical observations that were based on these results. These calculations were used 
in a search for CD$_3$OH spectral features in data from the ALMA PILS survey of the deeply 
embedded protostar IRAS 16293$-$2422. We detected CD$_3$OH confidently in the data with 
a large number of clearly separated and well-reproduced lines. 
A CD$_3$OH excitation temperature of 225~K was derived from the best fit, 
yielding a column density of $3.1 \times 10^{16}$~cm$^{-2}$. A comparison of this result to the column density of the CH$_3$OH main isotopolog deduced from optically thin lines of CH$_3^{18}$OH yields a CD$_3$OH/CH$_3$OH ratio as high as $\sim$0.33\%, implying that the triply deuterated variant is significantly enhanced compared to the D/H ratio inferred by comparing the singly and non-deuterated variants.


\begin{acknowledgements}
We acknowledge support by the Deutsche Forschungsgemeinschaft via the collaborative 
research center SFB~956 (project ID 184018867) project B3 as well as 
the Ger{\"a}tezentrum SCHL~341/15-1 (``Cologne Center for Terahertz Spectroscopy''). 
The research in Kharkiv and Braunschweig was carried out under support of the Volkswagen foundation. 
The assistance of the Science and Technology Center in the Ukraine is acknowledged 
(STCU partner project P756). J.K.J. is supported by the Independent Research Fund 
Denmark (grant number 0135-00123B). R.M.L. received support from the Natural 
Sciences and Engineering Research Council of Canada.
Our research benefited from NASA's Astrophysics Data System (ADS).
\end{acknowledgements}


\bibliographystyle{aa} 
\bibliography{CD3OH}

\newpage
\begin{table*}[]
\caption{\label{tbl:statisticInf} Overview of the dataset and the fit quality }
\begin{tabular}{lrl|lrl}
\hline
\multicolumn{3}{c|}{By measurement uncertainty}& \multicolumn{3}{c}{By torsional state} \\ 
\cline{1-6}
 
\multicolumn{1}{c}{Unc.$^a$} & \multicolumn{1}{c}{$\#^b$} & \multicolumn{1}{c|}{rms$^c$} 
& \multicolumn{1}{c}{$\varv_{\rm t}^d$} & \multicolumn{1}{c}{$\#^b$} & \multicolumn{1}{c}{wrms$^e$} \\

\cline{1-6}

0.010~MHz & 3183 & 0.0095~MHz & $\varv_{\rm t}=0 \leftarrow 0$ & 5371 & 0.80 \\
0.020~MHz &   50 & 0.0196~MHz & $\varv_{\rm t}=1 \leftarrow 1$ & 3105 & 0.99 \\
0.030~MHz & 2809 & 0.0237~MHz & $\varv_{\rm t}=2 \leftarrow 2$ &  104 & 0.94 \\
0.050~MHz &   77 & 0.0487~MHz & $\varv_{\rm t}=1 \leftarrow 0$ & 6137 & 0.68 \\
0.100~MHz &  828 & 0.0781~MHz & $\varv_{\rm t}=2 \leftarrow 1$ &  733 & 0.83 \\
0.200~MHz &  325 & 0.1935~MHz & $\varv_{\rm t}=2 \leftarrow 0$ &  348 & 0.79 \\
$4\times 10^{-4}$~cm$^{-1}$ & 5619 & $2.9\times 10^{-4}$~cm$^{-1}$ &   &  & \\ \hline

\end{tabular}
\tablefoot{$^{a}$ Estimated measurement uncertainties for each data group. $^{b}$ Number of lines (left part) or transitions (right part) of each category in the least-squares fit. Note that due to blending 15798  transitions correspond to 12891  measured line frequencies in the fit. $^{c}$ Root-mean-square (rms) deviation of corresponding data group. $^{d}$ Upper and lower state torsional quantum number $\varv_{\rm t}$. $^{e}$ Weighted root-mean-square (wrms) deviation of corresponding data group.}
\end{table*}



\newpage
\onecolumn
\begin{longtable}{lllc}
\caption{\label{tbl:ParametersTable} Fitted parameters of the RAM Hamiltonian for the CD$_3$OH molecule}\\

\hline\hline
$n_{tr}$\textit{$^a$} & Operator\textit{$^b$} & Par.\textit{$^{c}$} & Value\textit{$^{d,e}$} \\
\hline
\endfirsthead
\caption{continued.}\\
\hline\hline
$n_{tr}$\textit{$^a$} & Operator\textit{$^b$} & Par.\textit{$^{c}$} & Value\textit{$^{d,e}$} \\
\hline
\endhead
\hline

 $2_{2,0}$ & $p_\alpha^2$                & $F$         & $24.99356109(43)$\\
 $2_{2,0}$ & $(1-\cos 3\alpha)$          & $(1/2)V_3$  & $185.1289748(35)$  \\
 $2_{1,1}$ & $p_\alpha P_a$              & $\rho$      & $0.8946581266(13)$  \\
 $2_{0,2}$ & $P_a^2$                     & $A$  & $2.362252(81)$  \\
 $2_{0,2}$ & $P_b^2$                     & $B$  & $0.66232398(79)$  \\
 $2_{0,2}$ & $P_c^2$                     & $C$  & $0.64264483(79)$  \\
 $2_{0,2}$ & $(1/2)\{P_a{,}P_b\}$        & $2D_{ab}$   & $-0.008425955(88)$  \\
 $4_{4,0}$ & $(1-\cos 6\alpha)$          & $(1/2)V_6$ & $-1.4646101(58)$  \\ 
 $4_{4,0}$ & $p_\alpha ^4$               & $F_m$      & $-0.9213472(38)\times 10^{-2}$  \\
 $4_{3,1}$ & $p_\alpha ^3 P_a$           & $\rho_m$   & $-0.3632688(14)\times 10^{-1}$  \\
 $4_{2,2}$ & $P^2(1-\cos 3\alpha)$       & $V_{3J}$ & $-0.18368380(18)\times 10^{-2}$ \\
 $4_{2,2}$ & $P_a^2(1-\cos 3\alpha)$     & $V_{3K}$ & $0.57101(48)\times 10^{-2}$  \\ 
 $4_{2,2}$ & $(P_b^2-P_c^2)(1-\cos 3\alpha)$ & $V_{3bc}$ & $0.6587(36)\times 10^{-4}$  \\
 $4_{2,2}$ & $(1/2)\{P_a{,}P_b\}(1-\cos 3\alpha)$ & $V_{3ab}$ & $0.1395550(16)\times 10^{-1}$  \\  
 $4_{2,2}$ & $p^2_{\alpha} P^2$          & $F_J$ & $-0.7842383(16)\times 10^{-4}$  \\
 $4_{2,2}$ & $p^2_{\alpha} P_a^2$        & $F_K$ & $-0.5383863(18)\times 10^{-1}$ \\  
 $4_{2,2}$ & $p^2_{\alpha}(P_b^2-P_c^2)$ & $F_{bc}$ & $-0.187898(45)\times 10^{-3}$  \\
 $4_{2,2}$ & $(1/2)\{P_a{,}P_c\}\sin 3\alpha$ & $D_{3ac}$ & $0.177854(30)\times 10^{-1}$ \\
 $4_{2,2}$ & $(1/2)\{P_b{,}P_c\}\sin 3\alpha$ & $D_{3bc}$ & $-0.11072(10)\times 10^{-2}$ \\  
 $4_{1,3}$ & $p_\alpha P_aP^2$           & $\rho_J$ & $-0.12995447(30)\times 10^{-3}$ \\
 $4_{1,3}$ & $p_\alpha P_a^3$            & $\rho_K$ & $-0.3547761(11)\times 10^{-1}$ \\
 $4_{1,3}$ & $(1/2)\{P_a{,}(P_b^2-P_c^2)\}p_\alpha$ & $\rho_{bc}$ & $-0.258198(50)\times 10^{-3}$ \\
 $4_{0,4}$ & $P^4$                       & $-\Delta_J$ & $-0.7944(33)\times 10^{-6}$ \\
 $4_{0,4}$ & $P^2 P_a^2$                 & $-\Delta_{JK}$ & $-0.59151(14)\times 10^{-4}$  \\
 $4_{0,4}$ & $P_a^4$                     & $-\Delta_K$ & $-0.877157(12)\times 10^{-2}$ \\
 $4_{0,4}$ & $P^2(P_b^2-P_c^2)$          & $-2\delta_J$ & $-0.54786(13)\times 10^{-7}$  \\
 $4_{0,4}$ & $(1/2)\{P_a^2{,}(P_b^2-P_c^2)\}$ &  $-2\delta_K$ & $-0.73779(20)\times 10^{-4}$ \\
 $4_{0,4}$ & $(1/2)P^2\{P_a{,}P_b\}$     & $D_{abJ}$   & $-0.56434(79)\times 10^{-6}$  \\
 $4_{0,4}$ & $(1/2)\{P_a^3{,}P_b\}$      & $D_{abK}$   & $0.8214(11)\times 10^{-6}$  \\
 $6_{6,0}$ & $(1-\cos 9\alpha)$          & $(1/2)V_9$ & $2.1981^{fixed}$ \\   
 $6_{6,0}$ & $p_\alpha ^6$               & $F_{mm}$ & $0.62823(42)\times 10^{-5}$ \\
 $6_{5,1}$ & $p_\alpha ^5 P_a$           & $\rho_{mm}$ & $0.44437(23)\times 10^{-4}$ \\
 $6_{4,2}$ & $P^2(1-\cos 6\alpha)$       & $V_{6J}$ & $0.49894(39)\times 10^{-5}$ \\         
 $6_{4,2}$ & $P_a^2(1-\cos 6\alpha)$     & $V_{6K}$ & $0.673(30)\times 10^{-3}$ \\ 
 $6_{4,2}$ & $(P_b^2-P_c^2)(1-\cos 6\alpha)$ & $V_{6bc}$ & $-0.446(11)\times 10^{-4}$ \\
 $6_{4,2}$ & $(1/2)\{P_a{,}P_c\}\sin 6\alpha$  & $D_{6ac}$ & $0.1136(17)\times 10^{-2}$ \\     
 $6_{4,2}$ & $(1/2)\{P_b{,}P_c\}\sin 6\alpha$  & $D_{6bc}$ & $0.378(22)\times 10^{-4}$ \\     
 $6_{4,2}$ & $p_\alpha ^4P^2$            & $F_{mJ}$ & $0.48485(19)\times 10^{-7}$ \\    
 $6_{4,2}$ & $p_\alpha ^4P_a^2$          & $F_{mK}$ & $0.125829(52)\times 10^{-3}$ \\ 
 $6_{3,3}$ & $p_\alpha ^3P_aP^2$         & $\rho_{mJ}$ & $0.188581(67)\times 10^{-6}$ \\   
 $6_{3,3}$ & $p_\alpha ^3P_a^3$          & $\rho_{mK}$ & $0.184918(64)\times 10^{-3}$ \\               
 $6_{2,4}$ & $P^4(1-\cos 3\alpha)$       & $V_{3JJ}$ & $0.60397(57)\times 10^{-8}$ \\
 $6_{2,4}$ & $P^2P_a^2(1-\cos 3\alpha)$   & $V_{3JK}$ & $-0.16548(79)\times 10^{-6}$ \\
 $6_{2,4}$ & $P_a^4(1-\cos 3\alpha)$     & $V_{3KK}$ & $0.3187(56)\times 10^{-6}$ \\
 $6_{2,4}$ & $(1/2)P^2\{P_a{,}P_b\}(1-\cos 3\alpha)$ & $V_{3abJ}$ & $0.34485(79)\times 10^{-6}$ \\  
 $6_{2,4}$ & $(1/2)\{P_a^3{,}P_b\}(1-\cos 3\alpha)$ & $V_{3abK}$ & $-0.9540(14)\times 10^{-6}$ \\
 $6_{2,4}$ & $(1/2)\{P_a{,}P_b^3\}\cos 3\alpha$ & $V_{3ab3}$ & $0.5448(10)\times 10^{-6}$  \\  
 $6_{2,4}$ & $(1/2)\{P_b^2{,}P_c^2\}\cos 3\alpha$ & $V_{3b2c2}$ & $0.3925(12)\times 10^{-8}$  \\  
 $6_{2,4}$ & $p^2_{\alpha} P^4$          & $F_{JJ}$ & $0.48947(24)\times 10^{-9}$ \\       
 $6_{2,4}$ & $p^2_{\alpha} P_a^2 P^2$    & $F_{JK}$ & $0.278281(93)\times 10^{-6}$ \\       
 $6_{2,4}$ & $p^2_{\alpha} P_a^4$        & $F_{KK}$ & $0.149895(46)\times 10^{-3}$ \\ 
 $6_{2,4}$ & $(1/2)p^2_\alpha \{P_a^2{,}(P_b^2-P_c^2)\}$ & $F_{bcK}$ & $-0.642(13)\times 10^{-8}$ \\
 $6_{2,4}$ & $(1/2)P^2\{P_a{,}P_c\}\sin 3\alpha$ & $D_{3acJ}$ & $-0.13489(32)\times 10^{-6}$  \\  
 $6_{2,4}$ & $(1/2)\{P_a^3{,}P_c\}\sin 3\alpha$ & $D_{3acK}$ & $-0.3435(18)\times 10^{-6}$  \\  
 $6_{2,4}$ & $(1/2)\{P_a^2{,}P_b{,}P_c\}\sin 3\alpha$ & $D_{3bcK}$ & $-0.1279(16)\times 10^{-6}$  \\  
 $6_{2,4}$ & $(1/2)\{P_a{,}P_b^2{,}P_c\}\sin 3\alpha$ & $D_{3acb2}$ & $-0.56143(76)\times 10^{-6}$ \\
 $6_{1,5}$ & $p_\alpha P_aP^4$           & $\rho_{JJ}$ & $0.50052(35)\times 10^{-9}$ \\
 $6_{1,5}$ & $p_\alpha P_a^3 P^2$        & $\rho_{JK}$ & $0.184028(59)\times 10^{-6}$ \\
 $6_{1,5}$ & $p_\alpha P_a^5$            & $\rho_{KK}$ & $0.63858(18)\times 10^{-4}$ \\
 $6_{1,5}$ & $(1/2)P^2\{P_a{,}(P_b^2-P_c^2)\}p_\alpha$ & $\rho_{bcJ}$ & $0.6671(39)\times 10^{-9}$ \\
 $6_{1,5}$ & $(1/2)P^2\{P_a^2{,}P_b\}p_\alpha$  & $\rho_{abJ}$ & $-0.1092(16)\times 10^{-8}$ \\
 $6_{1,5}$ & $(1/2)\{P_a{,}P_b^2{,}P_c^2\}p_\alpha$ & $\rho_{b2c2}$ & $0.55339(72)\times 10^{-9}$  \\  
 $6_{0,6}$ & $P^6$                       & $\Phi_J$ & $0.2401(11)\times 10^{-12}$ \\
 $6_{0,6}$ & $P^4 P_a^2$                 & $\Phi_{JK}$ & $0.17666(14)\times 10^{-9}$ \\
 $6_{0,6}$ & $P^2P_a^4$                  & $\Phi_{KJ}$ & $0.45755(15)\times 10^{-7}$  \\    
 $6_{0,6}$ & $P_a^6$                     & $\Phi_K$ & $0.112081(29)\times 10^{-4}$ \\
 $6_{0,6}$ & $P^4(P_b^2-P_c^2)$         & $2\phi_J$ & $0.14279(30)\times 10^{-12}$ \\ 
 $6_{0,6}$ & $(1/2)P^2\{P_a^2{,}(P_b^2-P_c^2)\}$ & $2\phi_{JK}$ & $0.6740(37)\times 10^{-9}$ \\
 $6_{0,6}$ & $(1/2)\{P_a^4{,}(P_b^2-P_c^2)\}$ & $2\phi_K$ & $0.611(13)\times 10^{-8}$ \\
 $6_{0,6}$ & $(1/2)P^2\{P_a^3{,}P_b\}$      & $D_{abJK}$   & $-0.1162(15)\times 10^{-8}$  \\
 $8_{6,2}$ & $P_a^2(1-\cos 9\alpha)$      & $V_{9K}$ & $-0.269(11)\times 10^{-2}$  \\  
 $8_{6,2}$ & $(1/2)\{P_a{,}P_c\}\sin 9\alpha$ & $D_{9ac}$ & $-0.2699(55)\times 10^{-2}$ \\
 $8_{6,2}$ & $p_\alpha ^6P_a^2$          & $F_{mmK}$ & $0.681(25)\times 10^{-13}$ \\ 
 $8_{4,4}$ & $P^2 P_a^2(1-\cos 6\alpha)$ & $V_{6JK}$ & $-0.1478(49)\times 10^{-6}$ \\     
 $8_{4,4}$ & $P_a^4(1-\cos 6\alpha)$     & $V_{6KK}$ & $-0.1176(36)\times 10^{-5}$ \\
 $8_{4,4}$ & $P^2(P_b^2-P_c^2)(1-\cos 6\alpha)$ & $V_{6bcJ}$ & $0.1847(14)\times 10^{-8}$ \\    
 $8_{4,4}$ & $(1/2)\{P_a{,}P_b^3\}\cos 3\alpha$ & $V_{6ab3}$ & $0.9144(75)\times 10^{-7}$  \\  
 $8_{4,4}$ & $P^4p_\alpha ^4$          & $F_{mJJ}$ & $0.4298(95)\times 10^{-13}$ \\ 
 $8_{4,4}$ & $(1/2)P^2\{P_a{,}P_c\}\sin 6\alpha$ & $D_{6acJ}$ & $-0.1862(46)\times 10^{-7}$  \\ 
 $8_{4,4}$ & $(1/2)P^2\{P_b{,}P_c\}\sin 6\alpha$ & $D_{6bcJ}$ & $-0.2038(26)\times 10^{-8}$  \\ 
 $8_{4,4}$ & $(1/2)\{P_a^2{,}P_b{,}P_c\}\sin 6\alpha$ & $D_{6bcK}$ & $0.3900(54)\times 10^{-7}$  \\ 
 $8_{4,4}$ & $(1/2)\{P_a{,}P_b^2{,}P_c\}\sin 6\alpha$ & $D_{6acb2}$ & $0.2581(22)\times 10^{-6}$ \\
 $8_{4,4}$ & $(1/2)(\{P_b{,}P_c^3\}-\{P_b^3{,}P_c\})\sin 6\alpha$ & $D_{6bcbc}$ & $0.4782(31)\times 10^{-8}$\\ 
 $8_{2,6}$ & $P^6(1-\cos 3\alpha)$       & $V_{3JJJ}$ & $-0.3589(82)\times 10^{-13}$ \\
 $8_{2,6}$ & $P^4P_a^2(1-\cos 3\alpha)$   & $V_{3JJK}$ & $0.2083(16)\times 10^{-11}$ \\
 $8_{2,6}$ & $P_a^6(1-\cos 3\alpha)$   & $V_{3KKK}$ & $0.1868(15)\times 10^{-9}$ \\
 $8_{2,6}$ & $(1/2)P^2\{P_a^3{,}P_b\}(1-\cos 3\alpha)$ & $V_{3abJK}$ & $0.947(30)\times 10^{-10}$ \\  
 $8_{2,6}$ & $(1/2)\{P_a^5{,}P_b\}\cos 3\alpha$ & $V_{3abKK}$ & $0.1404(31)\times 10^{-9}$ \\
 $8_{2,6}$ & $P^4(P_b^2-P_c^2)(1-\cos 3\alpha)$ & $V_{3bcJJ}$ & $-0.5997(82)\times 10^{-13}$ \\
 $8_{2,6}$ & $(1/2)\{P_a^3{,}P_b^3\}\cos 3\alpha$ & $V_{3ab3K}$ & $0.1597(40)\times 10^{-9}$  \\  
 $8_{2,6}$ & $(1/2)P^2\{P_a^3{,}P_c\}\sin 3\alpha$ & $D_{3acJK}$ & $0.3041(26)\times 10^{-9}$  \\  
 $8_{2,6}$ & $(1/2)\{P_a^5{,}P_c\}\sin 3\alpha$ & $D_{3acKK}$ & $-0.3303(66)\times 10^{-9}$  \\  
 $8_{2,6}$ & $(1/2)\{P_a^4{,}P_b{,}P_c\}\sin 3\alpha$ & $D_{3bcKK}$ & $-0.1690(29)\times 10^{-9}$  \\  
 $8_{2,6}$ & $(1/2)P^2\{P_a{,}P_b^2{,}P_c\}\sin 3\alpha$ & $D_{3acb2J}$ & $0.515(26)\times 10^{-11}$ \\
 $8_{2,6}$ & $(1/2)\{P_a^3{,}P_b^2{,}P_c\}\sin 3\alpha$ & $D_{3acb2K}$ & $-0.11050(85)\times 10^{-8}$ \\
 $8_{2,6}$ & $(1/2)\{P_a{,}P_b^4{,}P_c\}\sin 3\alpha$ & $D_{3acb4}$ & $0.751(33)\times 10^{-11}$ \\
 $8_{2,6}$ & $(1/2)\{P_b^3{,}P_c^3\}\sin 3\alpha$ & $D_{3b3c3}$ & $0.7895(50)\times 10^{-12}$ \\ 
 $8_{2,6}$ & $p^2_{\alpha} P^6$        & $F_{JJJ}$ & $0.1247(15)\times 10^{-14}$ \\ 
 $8_{2,6}$ & $p^2_{\alpha} P^4P_a^2$        & $F_{JJK}$ & $-0.1329(24)\times 10^{-12}$ \\ 
 $8_{2,6}$ & $p^2_{\alpha} P^2P_a^4$        & $F_{JKK}$ & $-0.4239(74)\times 10^{-13}$ \\ 
 $8_{2,6}$ & $P^4(P_b^2-P_c^2)p^2_{\alpha}$ & $F_{bcJJ}$ & $-0.481(19)\times 10^{-15}$  \\
 $8_{1,7}$ & $P^4P_a^3 p_\alpha $  & $\rho_{JJK}$ & $-0.1032(14)\times 10^{-12}$ \\
 $8_{0,8}$ & $P_a^8$                       & $L_K$ & $-0.236(14)\times 10^{-13}$ \\
 $10_{6,4}$ & $P^2P_a^2(1-\cos 9\alpha)$       & $V_{9JK}$ & $0.688(18)\times 10^{-6}$ \\  
 $10_{6,4}$ & $P_a^4(1-\cos 9\alpha)$       & $V_{9KK}$ & $0.451(14)\times 10^{-5}$ \\  
 $10_{6,4}$ & $(1/2)\{P_b^2{,}P_c^2\}\cos 9\alpha$ & $V_{9b2c2}$ & $-0.1929(30)\times 10^{-6}$  \\  
 $10_{6,4}$ & $(1/2)(\{P_b{,}P_c^3\}-\{P_b^3{,}P_c\})\sin 9\alpha$ & $D_{9bcbc}$ & $-0.939(16)\times 10^{-7}$\\ 
 $10_{6,4}$ & $(1/2)\{P_a{,}P_b^2{,}P_c\}\sin 9\alpha$ & $D_{9acb2}$ & $-0.1002(29)\times 10^{-6}$ \\
 $10_{6,4}$ & $(1/2)\{P_a^2{,}P_b{,}P_c{,}p_\alpha^4{,} \sin 3\alpha  \}  $ & $D_{3bcmmK}$ & $-0.367(23)\times 10^{-12}$  \\  
 $10_{6,4}$ & $(1/2)\{P_a^3{,}P_c{,}p_\alpha^2{,} \sin 6\alpha  \}  $ & $D_{6acmK}$ & $0.1826(42)\times 10^{-9}$  \\  
 $10_{4,6}$ & $P_a^6(1-\cos 6\alpha)$       & $V_{6KKK}$ & $-0.3958(44)\times 10^{-9}$ \\
 $10_{4,6}$ & $(1/2)P^2\{P_b^2{,}P_c^2\}\cos 6\alpha$ & $V_{6b2c2J}$ & $0.2947(66)\times 10^{-12}$  \\  
 $10_{4,6}$ & $(1/2)\{P_a^3{,}P_b^3\}\cos 6\alpha$ & $V_{6ab3K}$ & $-0.5058(85)\times 10^{-10}$  \\  
 $10_{4,6}$ & $(1/2)P^2\{P_a{,}P_b^2{,}P_c\}\sin 6\alpha$ & $D_{6acb2J}$ & $-0.877(21)\times 10^{-11}$ \\
 $10_{2,8}$ & $(1/2)P^4\{P_b^2{,}P_c^2\}\cos 3\alpha$ & $V_{3b2c2JJ}$ & $-0.266(11)\times 10^{-16}$  \\  
 $10_{2,8}$ & $p^2_{\alpha} P^8$        & $F_{JJJJ}$ & $-0.379(34)\times 10^{-19}$ \\ 
 $12_{8,4}$ & $P^4(1-\cos 12\alpha)$       & $V_{12JJ}$ & $-0.1888(33)\times 10^{-6}$ \\  

\hline
\hline

\end{longtable}

\tablefoot{$^{a}$ \textit{n=t+r}, where \textit{n} is the total order of the operator, \textit{t} is the order 
  of the torsional part and \textit{r} is the order of the rotational part, respectively. The ordering scheme 
  of \citet{Nakagawa:1987} is used. $^{b}$ $\lbrace A,B,C,D,E \rbrace = ABCDE+EDCBA$. 
  $\lbrace A,B,C,D \rbrace = ABCD+DCBA$. $\lbrace A,B,C \rbrace = ABC+CBA$. $\lbrace A,B \rbrace = AB+BA$. 
  The product of the operator in the second column of a given row and the parameter in the third column 
  of that row gives the term actually used in the torsion-rotation Hamiltonian of the program, except for 
  \textit{F}, $\rho$ and \textit{$A_{\rm RAM}$}, which occur in the Hamiltonian in the form 
  $F(p_a + \rho P_a)^2 + A_{\rm RAM}P_a^2$. $^{c}$ The parameter nomenclature is based on the subscript 
  procedure of \citet{XU:2008305}. $^{d}$ Values of the parameters in units of cm$^{-1}$, except for $\rho$, 
  which is unitless. $^{e}$ Statistical uncertainties are given in parentheses as one standard uncertainty 
  in units of the last digits.}


\begin{appendix}

\section{Fits to CD$_3$OH transitions in the ALMA data}

\begin{figure}
\resizebox{\hsize}{!}{\includegraphics{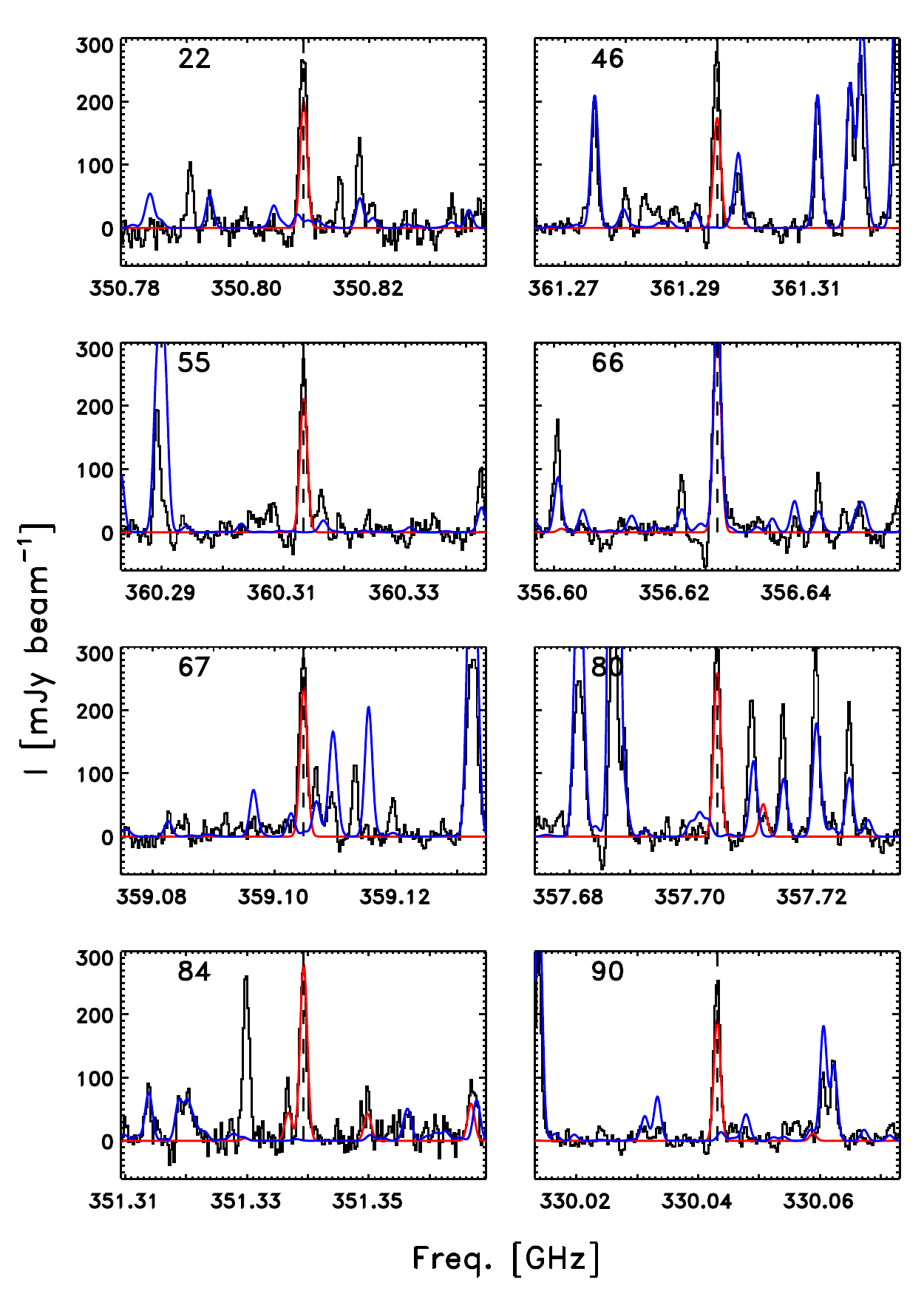}}
\caption{Fits to the 48 transitions of CD$_3$OH in the ALMA data predicted to be the strongest 
         while having $\tau < 0.2$ from the model described in Sect.~\ref{astrosearch}: 
         the model for CD$_3$OH is shown in red and the combined model for other species 
         identified in PILS in blue. The spectra are sorted according to the energy of 
         the upper level above the ground-state, indicated in the upper left corner of each plot.}
\label{all_spec1}
\end{figure}
\begin{figure}
\resizebox{\hsize}{!}{\includegraphics{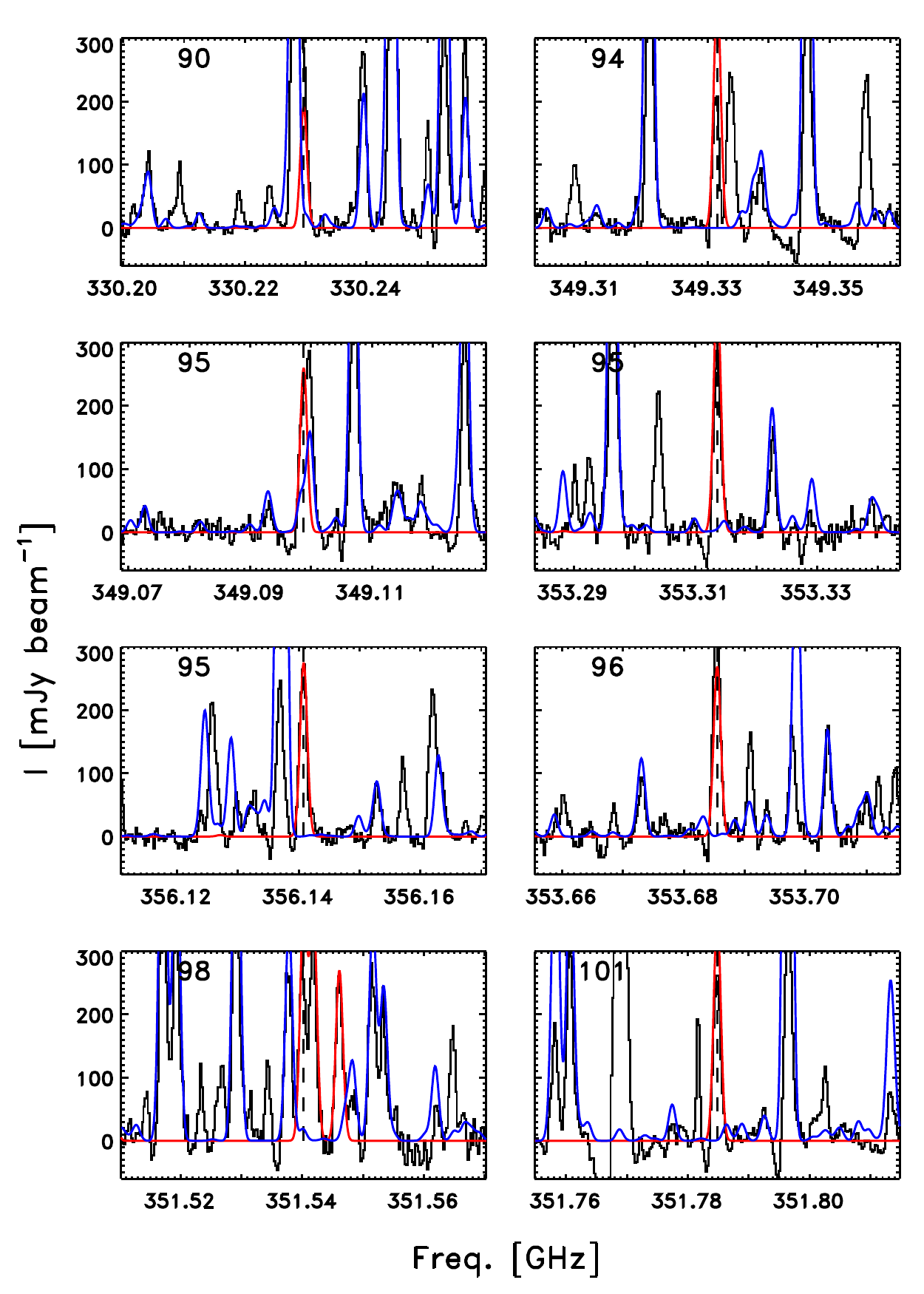}}
\caption{Fits to CD$_3$OH continued -- as in Fig.~\ref{all_spec1}.}
\end{figure}
\begin{figure}
\resizebox{\hsize}{!}{\includegraphics{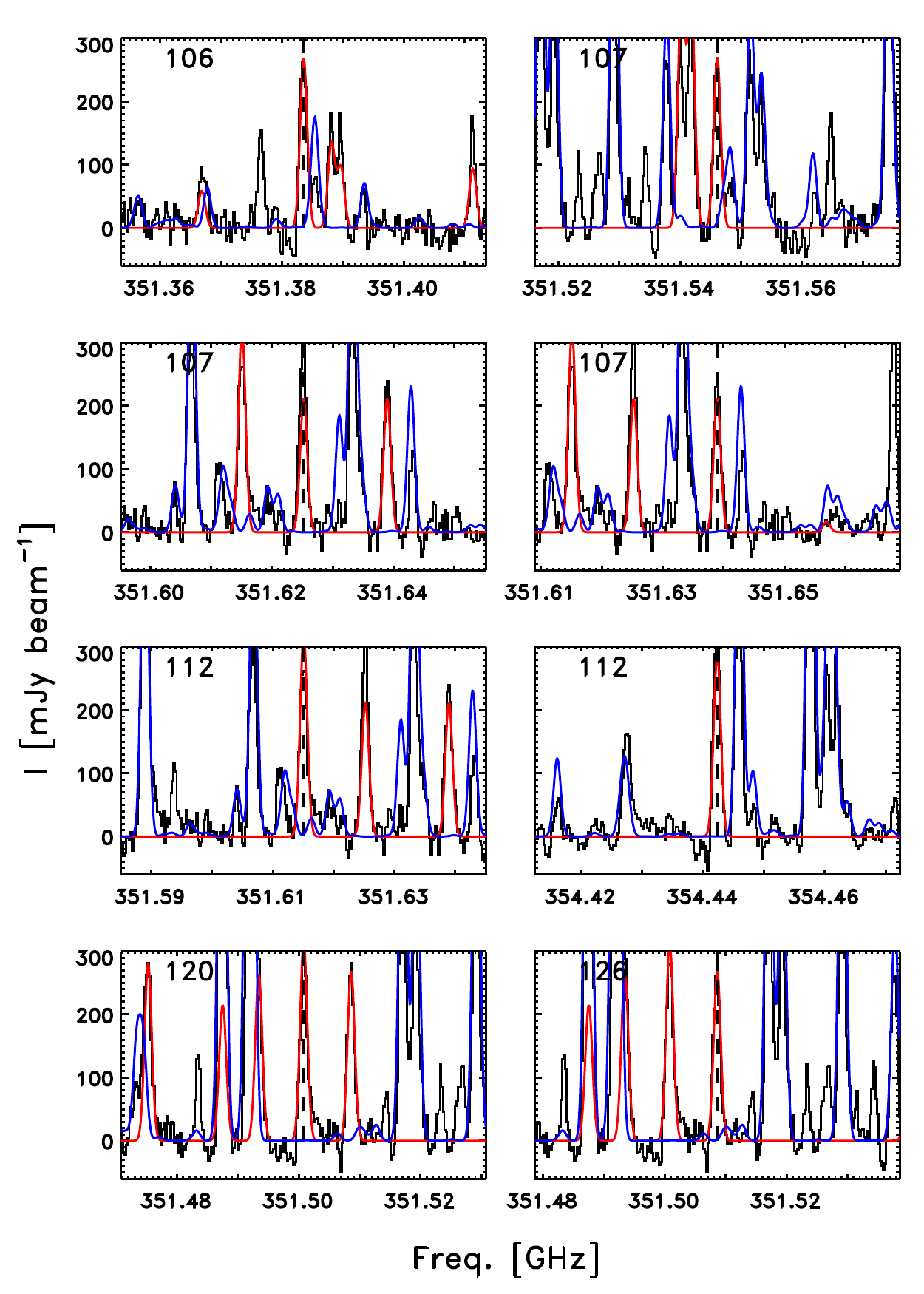}}
\caption{Fits to CD$_3$OH continued -- as in Fig.~\ref{all_spec1}.}
\end{figure}
\begin{figure}
\resizebox{\hsize}{!}{\includegraphics{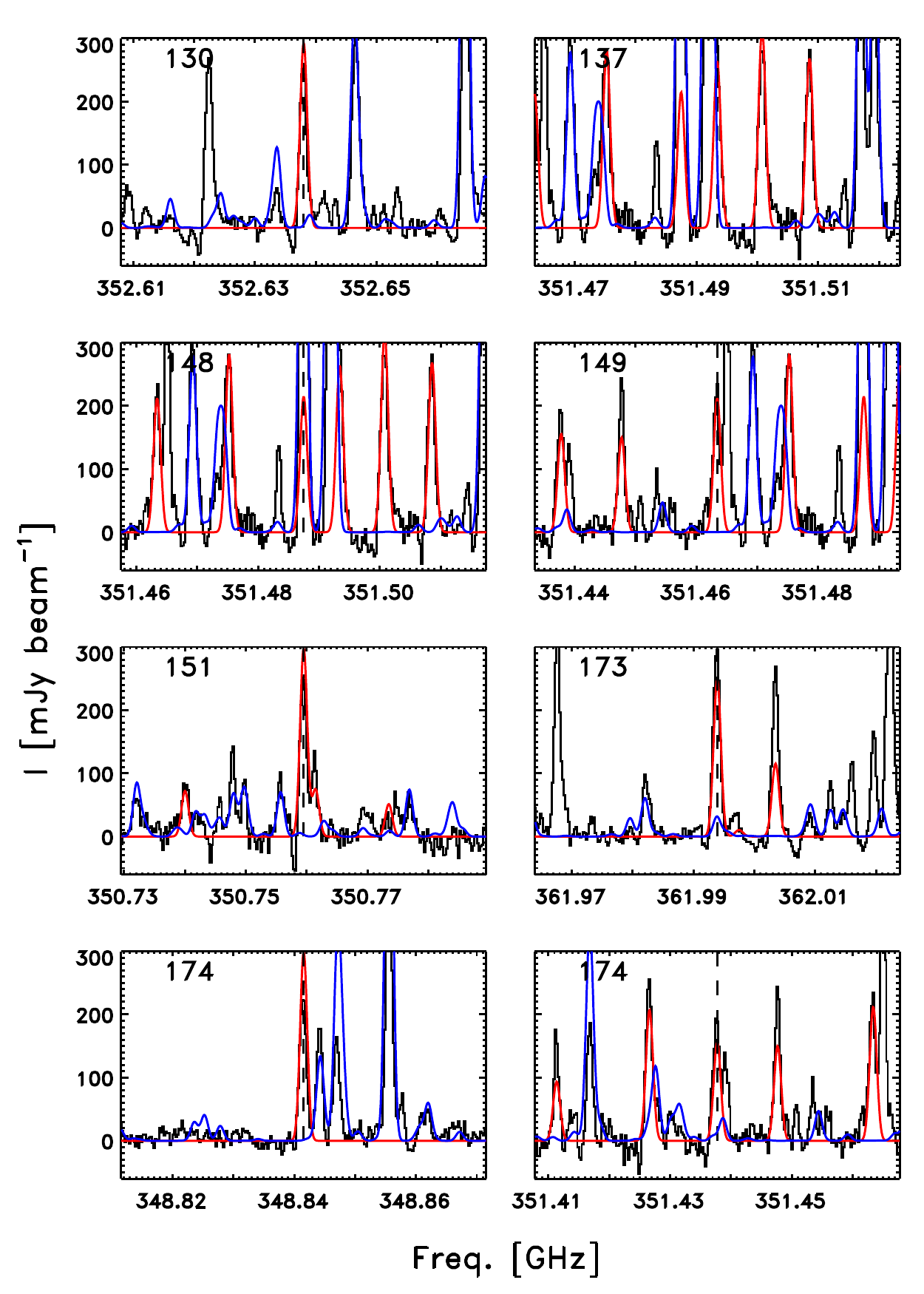}}
\caption{Fits to CD$_3$OH continued -- as in Fig.~\ref{all_spec1}.}
\end{figure}
\begin{figure}
\resizebox{\hsize}{!}{\includegraphics{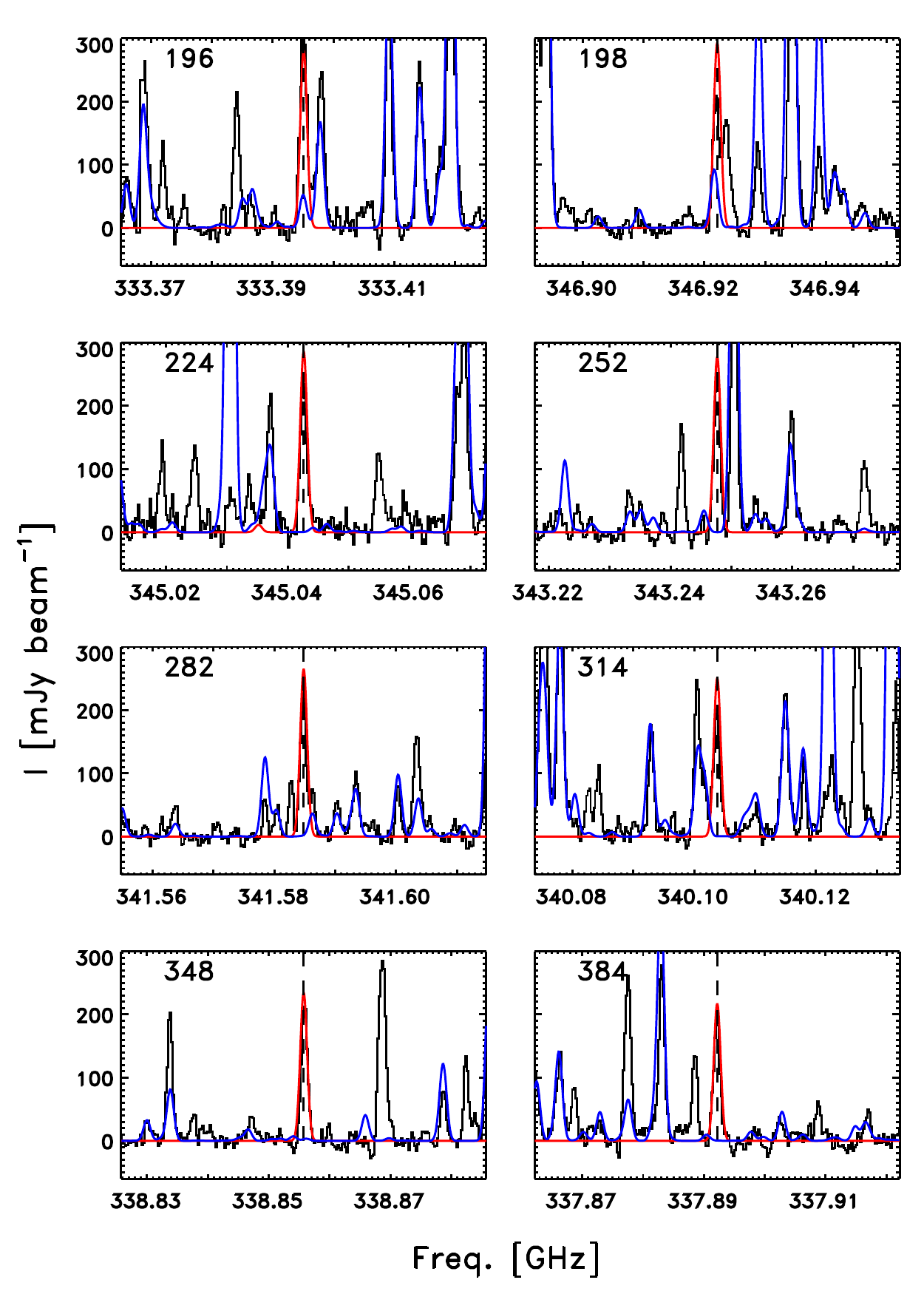}}
\caption{Fits to CD$_3$OH continued -- as in Fig.~\ref{all_spec1}.}
\end{figure}
\begin{figure}
\resizebox{\hsize}{!}{\includegraphics{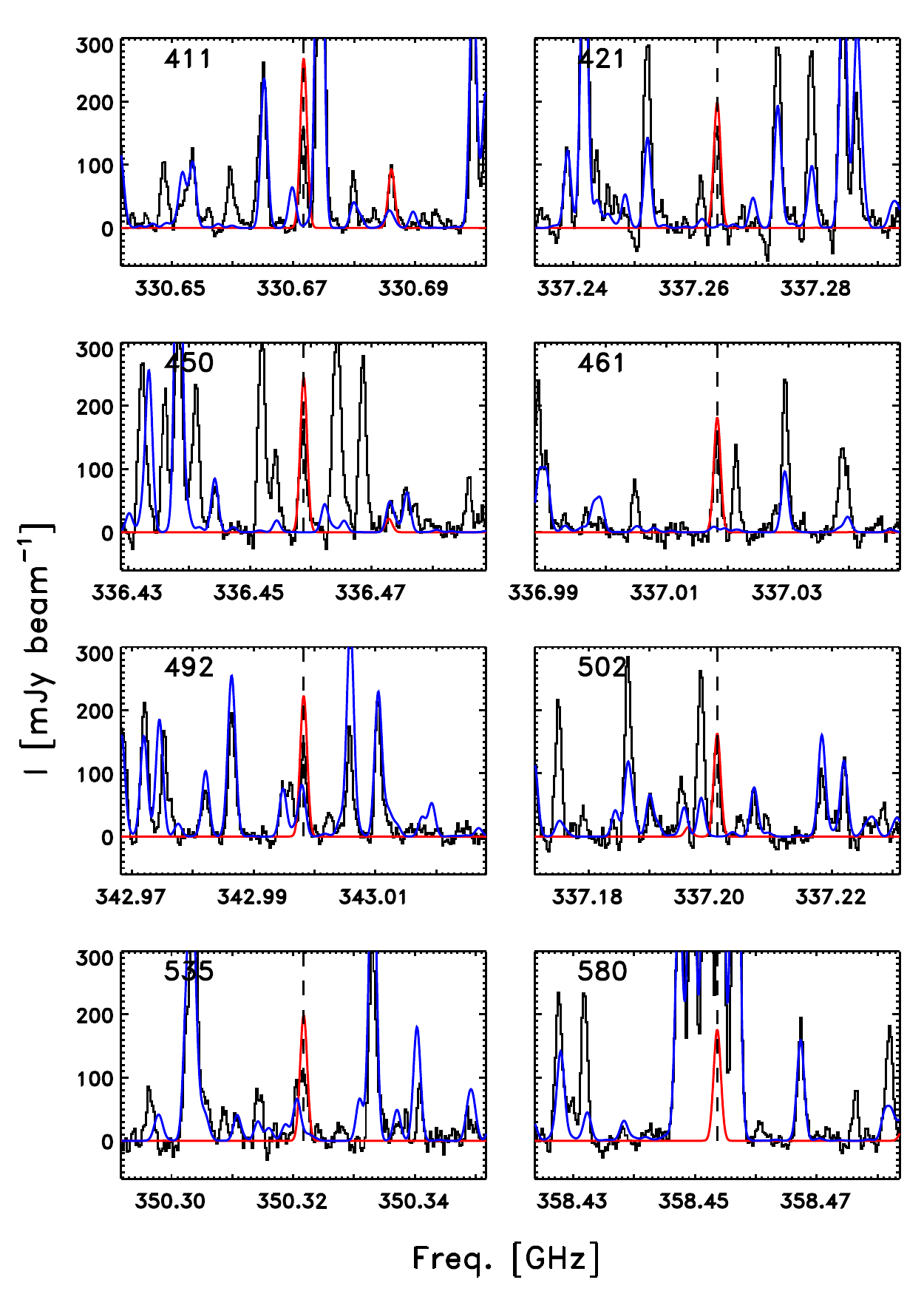}}
\caption{Fits to CD$_3$OH continued -- as in Fig.~\ref{all_spec1}.}\label{all_spec6}
\end{figure}

\begin{figure}
\resizebox{\hsize}{!}{\includegraphics{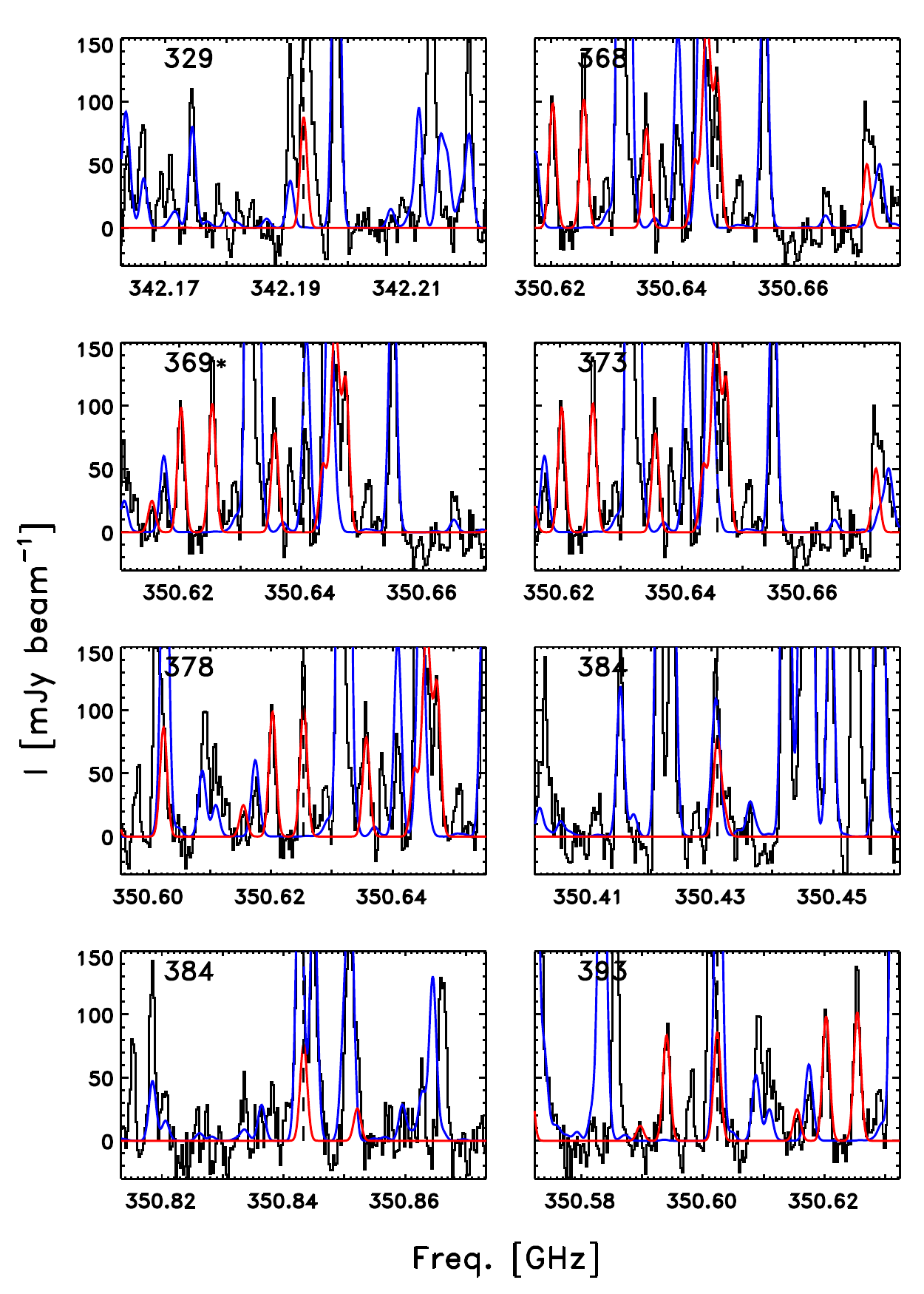}}
\caption{Fits to the $\varv_{\rm t}$ = 1  transitions of CD$_3$OH in the ALMA data predicted to be the strongest from the model described in Sect.~\ref{astrosearch}: 
         the model for CD$_3$OH is shown in red and the combined model for other species 
         identified in PILS in blue. The spectra are sorted according to the energy of 
         the upper level above the ground-state, indicated in the upper left corner of each plot.}
\label{all_spec7}
\end{figure}

\begin{figure}
\resizebox{\hsize}{!}{\includegraphics{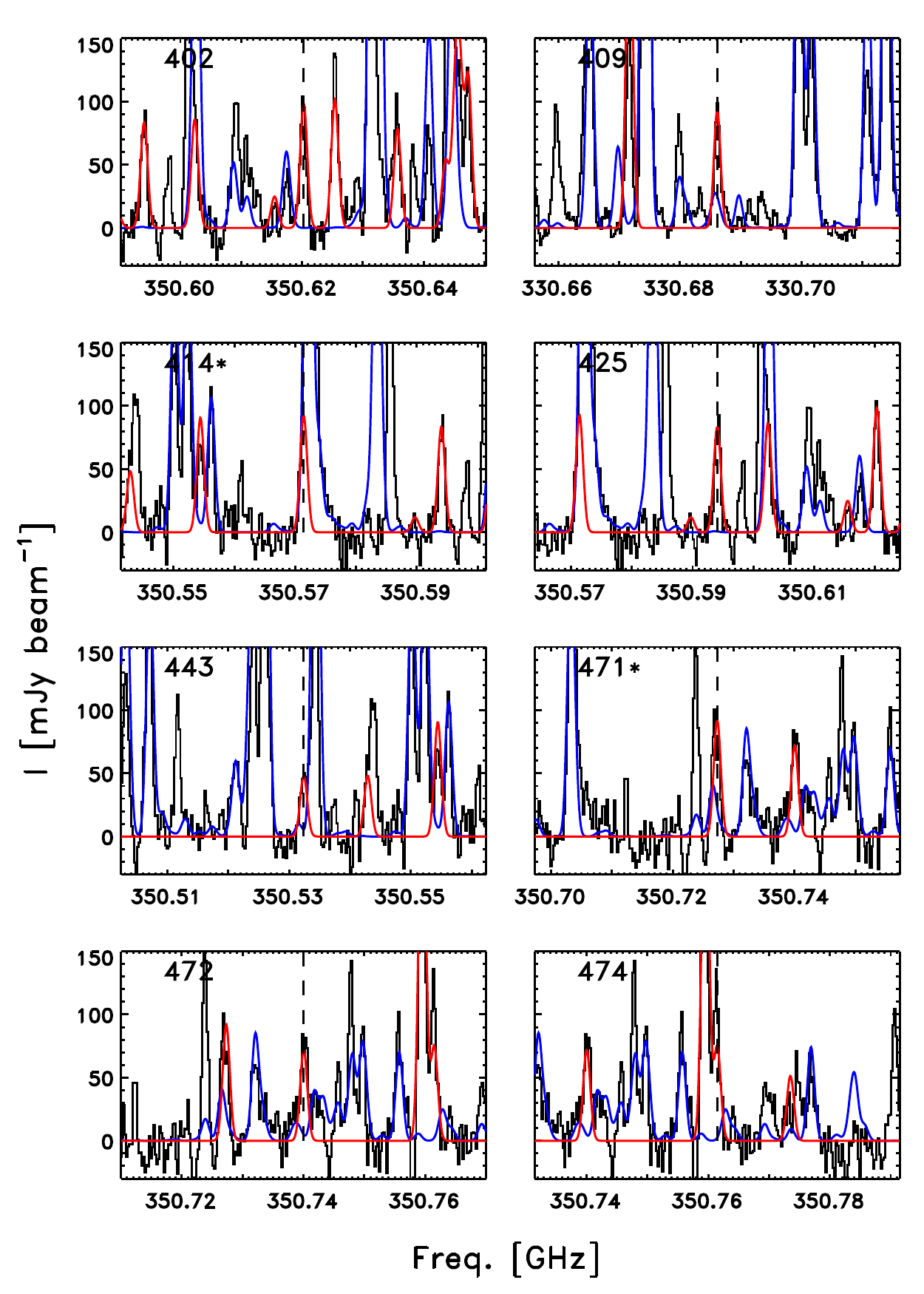}}
\caption{Fits to CD$_3$OH continued -- as in Fig.~\ref{all_spec7}.}\label{all_spec8}
\end{figure}

\begin{figure}
\resizebox{\hsize}{!}{\includegraphics{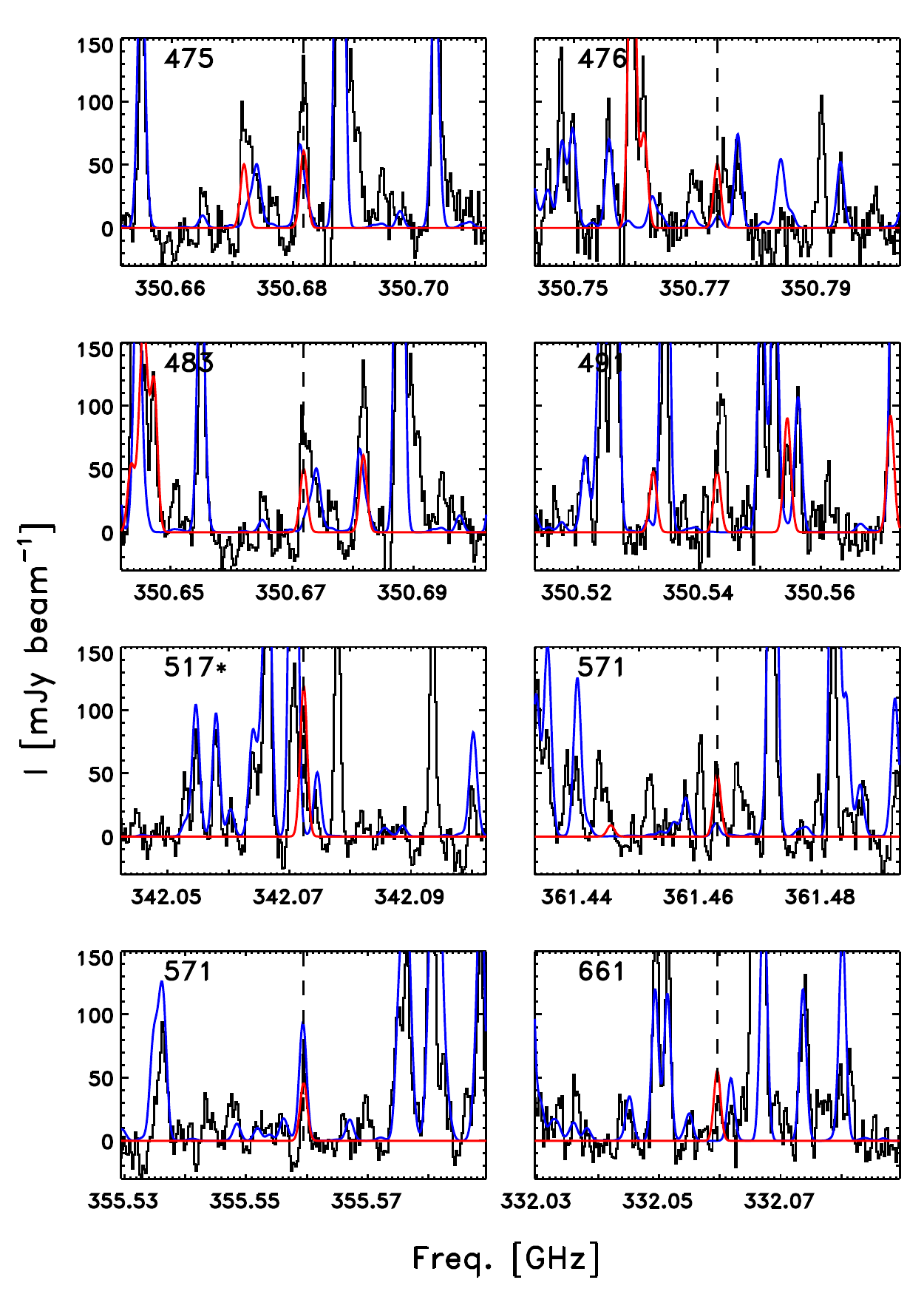}}
\caption{Fits to CD$_3$OH continued -- as in Fig.~\ref{all_spec7}.}\label{all_spec9}
\end{figure}

\end{appendix}

\end{document}